\def\arxiv{}
\def\fulltext{}
\def\figone{}

\documentclass[aps,prl,superscriptaddress,floatfix,nofootinbib,showkeys,showpacs,twocolumn,reprint]{revtex4-1}

\usepackage{aas_macros}
\usepackage{amsmath}
\usepackage{graphicx}
\usepackage[normalem]{ulem}

\newcommand{\shorten}[1]{\color{black}{#1}\color{black}}
\renewcommand{\shorten}[1]{}

\usepackage{color}
\newcommand{\upnum}{\textcolor{red}{\bf{XX}}}
\renewcommand{\upnum}{}
\newcommand{\jcap}{JCAP}
\newcommand{\nar}{New Astronomy Reviews}
\newcommand{\pass}{{\texttt{Pass~8}}}

\ifdefined\wc
\newcommand{\Eprint}[1]{ }
\fi

\begin{document}


\title{Search for Spectral Irregularities due to Photon--Axionlike-Particle Oscillations with the Fermi Large Area Telescope}


\author{M.~Ajello}
\affiliation{Department of Physics and Astronomy, Clemson University, Kinard Lab of Physics, Clemson, SC 29634-0978, USA}
\author{A.~Albert}
\affiliation{W. W. Hansen Experimental Physics Laboratory, Kavli Institute for Particle Astrophysics and Cosmology, Department of Physics and SLAC National Accelerator Laboratory, Stanford University, Stanford, CA 94305, USA}
\author{B.~Anderson}
\affiliation{Department of Physics, Stockholm University, AlbaNova, SE-106 91 Stockholm, Sweden}
\affiliation{The Oskar Klein Centre for Cosmoparticle Physics, AlbaNova, SE-106 91 Stockholm, Sweden}
\author{L.~Baldini}
\affiliation{Universit\`a di Pisa and Istituto Nazionale di Fisica Nucleare, Sezione di Pisa I-56127 Pisa, Italy}
\affiliation{W. W. Hansen Experimental Physics Laboratory, Kavli Institute for Particle Astrophysics and Cosmology, Department of Physics and SLAC National Accelerator Laboratory, Stanford University, Stanford, CA 94305, USA}
\author{G.~Barbiellini}
\affiliation{Istituto Nazionale di Fisica Nucleare, Sezione di Trieste, I-34127 Trieste, Italy}
\affiliation{Dipartimento di Fisica, Universit\`a di Trieste, I-34127 Trieste, Italy}
\author{D.~Bastieri}
\affiliation{Istituto Nazionale di Fisica Nucleare, Sezione di Padova, I-35131 Padova, Italy}
\affiliation{Dipartimento di Fisica e Astronomia ``G. Galilei'', Universit\`a di Padova, I-35131 Padova, Italy}
\author{R.~Bellazzini}
\affiliation{Istituto Nazionale di Fisica Nucleare, Sezione di Pisa, I-56127 Pisa, Italy}
\author{E.~Bissaldi}
\affiliation{Istituto Nazionale di Fisica Nucleare, Sezione di Bari, I-70126 Bari, Italy}
\author{R.~D.~Blandford}
\affiliation{W. W. Hansen Experimental Physics Laboratory, Kavli Institute for Particle Astrophysics and Cosmology, Department of Physics and SLAC National Accelerator Laboratory, Stanford University, Stanford, CA 94305, USA}
\author{E.~D.~Bloom}
\affiliation{W. W. Hansen Experimental Physics Laboratory, Kavli Institute for Particle Astrophysics and Cosmology, Department of Physics and SLAC National Accelerator Laboratory, Stanford University, Stanford, CA 94305, USA}
% ...

\date{\today}

\begin{abstract}
We report on the search for spectral irregularities induced by oscillations between photons 
and axionlike-particles (ALPs)
in the $\gamma$-ray spectrum of 
 NGC\,1275, the central galaxy of the Perseus cluster.
Using six years of \emph{Fermi} Large Area Telescope data, we find 
 no evidence for ALPs 
 and exclude couplings above $5\,\times10^{-12}\,\mathrm{GeV}^{-1}$ 
 for ALP masses $0.5 \lesssim m_a \lesssim 5$\,neV  at 95\,\% confidence.
The limits are 
competitive with the sensitivity of 
planned laboratory experiments, 
and,  
together with other bounds, 
strongly constrain the possibility that ALPs 
can reduce the $\gamma$-ray opacity of the Universe.
\end{abstract}

\pacs{}

\ifdefined\fulltext
\maketitle
\fi

\section{\label{sec:intro}Introduction}

Axions and axionlike-particles (ALPs)
are predicted by a variety of extensions of the Standard Model \cite{jaeckel2010,pq1977,weinberg1978,wilczek1978,witten1984,ringwald2014}.
If produced non-thermally in the early Universe, these particles may account for all or a significant fraction 
of the cold dark matter (DM) \cite[e.g.][]{preskill1983,abbott1983,marsh2011,arias2012},
and could be detected through their 
coupling to photons in
magnetic fields  \cite{sikivie1983,raffelt1988}.
While the axion mass is proportional to its coupling to photons,  
these two parameters are independent in the case of ALPs.

Photon-ALP interactions could leave an imprint on $\gamma$-ray spectra, provided that the ALP mass is sufficiently small,
$m_a \lesssim\,\mu$eV.
Above a critical energy $E_\mathrm{crit}$ 
photon-ALP mixing becomes maximal, leading to a reduction of the photon flux \cite{hooper2007:alps,hochmuth2007,deangelis2008}. 
Around $E_\mathrm{crit}$ this is accompanied by 
spectral irregularities
that depend on the strength and morphology of the 
magnetic field \cite{wouters2012}.
Photon-ALP conversions could also 
reduce the opacity of the Universe caused by pair production of $\gamma$ rays
with photons of the extragalactic background light (EBL) \cite{deangelis2007,simet2008}. 
Evidence exists that the $\gamma$-ray absorption is indeed lower than expected from state-of-the-art EBL models \cite{deangelis2009,deangelis2011,horns2012,rubtsov2014}, 
and ALPs have been used to explain these observations \cite{simet2008,sanchezconde2009,dominguez2011alps,meyer2013,galanti2015} 
(see, however, \cite{biteau2015,dominguez2015}).  

Sources embedded in galaxy clusters are promising to search for ALPs due to the strong  
magnetic fields extending over large spatial scales in these systems.  
For example, the absence of irregularities above 200\,GeV in the 
spectrum of the blazar PKS\,2155-304, 
associated with a poor galaxy cluster, has 
been used to constrain the photon-ALP coupling \cite{hess2013:alps}.
Here, we focus on 
the search for irregularities in the spectrum
 of the radio galaxy NGC\,1275 
with the \emph{Fermi} Large Area Telescope (LAT).
NGC\,1275 is the most favorable target since it is a bright $\gamma$-ray emitter detected with a significance exceeding $100\,\sigma$ 
in the third \emph{Fermi}-LAT source catalog (3FGL) \cite{3FGL}.
Its broadband emission can be explained with synchrotron-self Compton models, which predict a smooth $\gamma$-ray
 spectrum \cite{aleksic2014,tavecchio2014:ngc1275}. 
It is located at the center of the Perseus cool-core cluster for which
rotation measures (RMs) suggest a high central magnetic field \cite{taylor2006}.

Our analysis makes use of the newest {\pass} event-level analysis for LAT data.
Compared to previous Passes, {\pass} has an improved angular resolution,
 a broader energy range, larger effective area, as well as reduced uncertainties 
in the instrumental response functions (IRFs) \cite{atwood2013pass8}.

\section{LAT data selection}
We make use of six years of LAT data taken between Aug.\,4, 2008 and Aug.\,4, 2014 
in the energy range from 100\,MeV to 500\,GeV.
 For lower energies, the effective area decreases rapidly and the energy dispersion increases.
At energies above 500\,GeV we do not expect sufficient photon statistics \cite{2fhl}.
We only consider events that arrive at a zenith angle $\theta_z < 90^\circ$ in order to minimize 
the contribution of $\gamma$ rays from the Earth limb.
Time intervals that correspond to bright solar flares and $\gamma$-ray bursts are excluded.
We extract $\gamma$-ray like events within a $10^\circ\times10^\circ$ region of interest (ROI) 
 centered at 
the position of NGC\,1275: $\alpha_{2000} = 3^\mathrm{h}19^\mathrm{m}49.9^\mathrm{s}$, $\delta_{2000}=+41^\circ30^\mathrm{m}49.2^\mathrm{s}$ \cite{3FGL}.

Events passing the {\pass} \texttt{P8R2\_SOURCE} selection cuts are analyzed using the \texttt{P8R2\_SOURCE\_V6} IRFs.\footnote{\url{http://www.slac.stanford.edu/exp/glast/groups/canda/lat_Performance.htm}}
An innovation of the {\pass} IRFs is the possibility to subdivide an event class into 
event types according to the quality of the angular 
 or energy reconstruction (PSF and EDISP event types, respectively).
In this analysis we will use the EDISP types to maximize our sensitivity to spectral irregularities.
Events are classified into one of four types
 ranging from EDISP0 to EDISP3, that denote the quality of the energy reconstruction from worst to best.
All EDISP event types have a similar number of events in each logarithmic energy bin and are mutually exclusive.
The energy dispersion matrices are given in the Supplemental Material
\cite{supplements}.

\section{Photon-ALP oscillations}
Following \cite[e.g.][]{grossmann2002,csaki2003,mirizzi2008,deangelis2011,meyer2014}, 
we derive the probability $P_{\gamma\gamma}$ for a final state photon in the photon-ALP beam  
as a function of energy
 for an initially un-polarized 
 photon beam (see the Supplemental Material). 
We expect the irregularities to occur around a critical energy \cite{hooper2007:alps}, 
\ifdefined\fulltext
\begin{equation}
E_\mathrm{crit} \sim 
2.5\,\mathrm{GeV}\, \frac{|m^2_{a,\mathrm{neV}} - \omega_\mathrm{pl,\,neV}^2|}{g_{11} B_{\mu\mathrm{G}}},
\end{equation}
\fi
with ALP mass $m_{a,\mathrm{neV}}$ and 
plasma frequency $\omega_\mathrm{pl,\,neV}$ in units of neV, coupling constant $g_{11} = g_{a\gamma} / 10^{-11} \mathrm{GeV}^{-1}$, and 
magnetic field $B_{\mu\mathrm{G}} = B / 1\,\mu\mathrm{G}$.
We include photon-ALP mixing in the intra-cluster and Galactic magnetic fields \cite{simet2008,horns2012ICM}. 
The $B$ field of the Milky Way is modeled with the coherent component of the model
described in \cite{jansson2012}. We do not include its turbulent component, as the scales on which the turbulence occurs are usually smaller than the photon-ALP oscillation length. 
The turbulent intra-cluster $B$ field is described below.
Absorption of $\gamma$ rays by the EBL is taken into account through the model of \cite{dominguez2011}. 
We neglect any oscillations in the intergalactic magnetic field (IGMF). 
With current upper limits on the IGMF strength of $\lesssim10^{-9}\,$G and 
on the photon-ALP coupling, $g_{11} < 6.6$ \cite{ayala2014}, we find that 
$E_\mathrm{crit} \lesssim 100\,$GeV only for $m_{a,\mathrm{neV}} \lesssim 0.5$. 
For such low masses, $g_{11}$ is further constrained below $0.6$ 
from the non-observation of $\gamma$ rays from SN1987A \cite{payez2015}.
Given this small coupling and the comparatively short distance to NGC\,1275 
 (redshift $z = 0.017559$), no strong irregularities should be induced by mixing in the IGMF. 

\subsection{Intra-cluster magnetic field}
Faraday RM observations 
and magneto-hydro\-dy\-namic simulations suggest that 
the magnetic field in galaxy clusters is turbulent and that its strength follows 
the electron density $n_e(r)$ of the intra-cluster medium (ICM), $B(r) = B_0 (n_e(r) / n_e(r = 0))^\eta$ \cite{dolag2008,dubois2008,feretti2012}.
We model the turbulent component as a divergence-free homogeneous isotropic field with Gaussian turbulence 
with zero mean and a variance $\sigma_B$ \cite{meyer2014}. The energy density follows a power law $M(k) \propto k^q$ in wave numbers $k$. 
It is non-zero only between the minimum and maximum turbulence scales $k_L = 2\pi/\Lambda_\mathrm{max}$ and $k_H = 2\pi/\Lambda_\mathrm{min}$.

For the Perseus cluster, we use $n_e(r)$ derived from X-ray observations (Eq. (4) in \cite{churazov2003}) 
within the inner $r_\mathrm{max} = 500\,$kpc. Beyond this radius, we conservatively assume a zero magnetic field. 
RMs 
currently only probe the innermost region (tens of pc) around NGC\,1275. 
The observations lead to an estimated central magnetic field of $25\,\mu$G \cite{taylor2006}. 
An independent lower limit
of $B_0 \gtrsim 2 {-} 13 \,\mu$G for $0.3 \leq \eta \leq 0.7$
has been derived from 
MAGIC observations of the Perseus cluster 
 \cite{magic_perseus2012}.
These results motivate our assumptions for $\sigma_B = 10\,\mu$G and $\eta = 0.5$, which are also in line with observations of other cool-core clusters \cite[e.g.][]{kuchar2011,vacca2012}. 

For the turbulence spectrum, we assume values derived from RMs of the cool-core cluster A\,2199 \cite{vacca2012}, which has a comparable number 
of member galaxies.
The fiducial parameter choices are summarized in Tab. \ref{tab:model}. 

\begin{table}[thb]
\ifdefined\fulltext
\begin{tabular}{lc}
\hline\hline
Parameter & Value\\
\hline
$\sigma_B$ & $10\,\mu$G\\
$r_\mathrm{max}$ &  500\,kpc \\
$\eta$ & 0.5 \\
$q$ & $-2.8$ \\
$\Lambda_\mathrm{min}$ & 0.7\,kpc\\
$\Lambda_\mathrm{max}$ & 35\,kpc\\
\hline
\end{tabular}
\fi
\caption{Fiducial model parameters for the intra-cluster magnetic field in Perseus.}
\label{tab:model}
\end{table}

\section{Data Analysis}
We perform a binned Poisson likelihood analysis, similar to the DM signal search from dwarf spheroidal galaxies \cite{dsphPass7,dsphPass8}. 
Events are binned into $10^\circ\times10^\circ$ sky maps with a resolution of $0.2^\circ$ per pixel.
The width of the logarithmically spaced energy bins is chosen to be 30\,\% of the 
median energy resolution
 of each EDISP event type  
(see the Supplemental Material for details).
This results in 39, 67, 94, and 145 energy bins for EDISP0-3, respectively.
We have tested with simulations that bin sizes below 40\,\% of the median energy resolution 
do not affect the results. 

For each event type, 
we perform a fit over the entire energy range and ROI for all source parameters (nuisance parameters $\boldsymbol{\theta}_{i}$)
using  \emph{gtlike} included in the \emph{Fermi}-LAT \emph{Science Tools} version v10r01p01.\footnote{\url{http://fermi.gsfc.nasa.gov/ssc/data/analysis/}}
We include all point sources listed in the 3FGL within $15^\circ$ from the ROI center. 
The diffuse backgrounds are modeled with templates for the 
 Galactic and the isotropic extragalactic $\gamma$-ray emission.\footnote{\url{http://fermi.gsfc.nasa.gov/ssc/data/access/lat/BackgroundModels.html}}
The energy dispersion is taken into account  
in the fitting of the point sources
whereas it is already accounted for in the 
the data-driven derivation of the diffuse templates. 
Normalizations of the diffuse sources and point sources within $8^\circ$ 
from the ROI center are left free to vary. 
All spectral indices of the point sources within $4^\circ$ are also free parameters. 
The time-averaged spectrum of NGC\,1275 is modeled with a logarithmic (log) parabola,
 $F(E) = N (E/E_0)^{-(\alpha + \beta\ln(E/E_0))}$,
where $E_0$ is fixed to 530\,MeV \cite{3FGL}.

Under the assumption that the profiled nuisance parameters 
do not change when 
considering each bin separately \cite{dsphPass7}, 
we extract the likelihood in each reconstructed energy bin $k'$, 
$\mathcal{L}(\mu_{ik'}, \boldsymbol{\theta}_{i} | D_{ik'})$  as a function of expected counts $\mu_{ik'}$ of NGC\,1275, and observed counts $D_{ik'}$. 
For NGC\,1275 a power law with fixed spectral index $\Gamma = 2$ is now assumed in each bin. For each tested value of  $\mu_{ik'}$ we re-optimize the normalization of the spectrum of the radio galaxy IC\,310
which has an angular separation of $\sim0.6^\circ$ from NGC\,1275.

Under the ALP hypothesis, characterized by $P_{\gamma\gamma} \equiv 
P_{\gamma\gamma}(E, m_a, g_{a\gamma}, \mathbf{B}_j)$ for one random turbulent $B$-field realization $\mathbf{B}_j$,
the expected number of photons is calculated through
\ifdefined\fulltext
\begin{equation}
\mu_{ik'} =  \sum\limits_{k} \mathcal{D}^i_{kk'} \int\limits_{\Delta E_k} dE\, P_{\gamma\gamma} F(E) \mathcal{E}^i (E),\label{eq:expcounts}
\end{equation}
\fi
where the integration runs over the true energy bin $\Delta E_k$,  $\mathcal{E}^i$ is the exposure, and $\mathcal{D}^i_{kk'}$ is the energy dispersion 
 for event type EDISP$i$. 
 Under the null hypothesis, $P_{\gamma\gamma}$ reduces to the 
  EBL attenuation.
The parameters of the intrinsic source spectrum $F(E)$, $N,\alpha$, and $\beta$, 
are further nuisance parameters.
For each tested ALP parameter and magnetic field, we determine these parameters by profiling the joint likelihood of all energy bins $k'$
\ifdefined\fulltext
\begin{equation}
\mathcal{L}_i(\boldsymbol{\mu}, \boldsymbol{\theta} | \mathbf{D}) \equiv \prod\limits_{k'} \mathcal{L}(\mu_{ik'}, \boldsymbol{\theta}_{i} | D_{ik'}),
\end{equation}
\fi
for each event type separately, using the pre-computed likelihood curves $\mathcal{L}(\mu_{ik'}, \boldsymbol{\theta}_{i} | D_{ik'})$.
In this way, we treat each event type selection as an independent measurement.\footnote{
This procedure will result in different best-fit estimators for the source parameters for each event type. 
In this way, it is possible to speed up the optimization considerably. 
We have verified that our results do not change when the parameters of NGC\,1275 are tied 
over the event types.
}
The bin-by-bin likelihood curves for the EDISP3 event type are shown in Fig. \ref{fig:sed}
together with the best-fit spectra. 

\ifdefined\figone
\begin{figure}[thb]
\ifdefined\fulltext
\includegraphics[width = 1\linewidth]{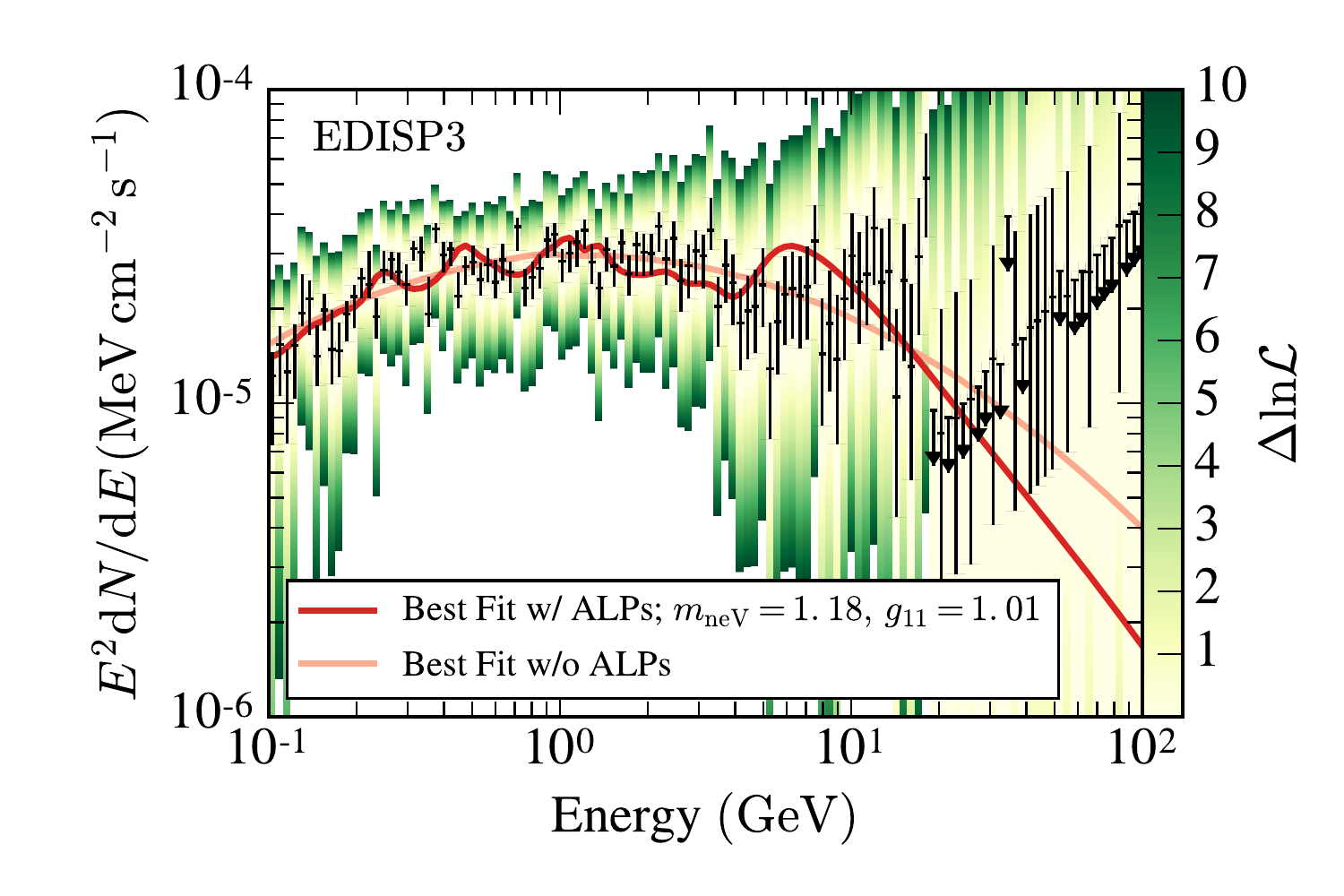}
\fi
\caption{
\label{fig:sed}
The 
 likelihood curves (shown in color) for the EDISP3 event type.
$\Delta\ln\mathcal{L} = 0$
corresponds to the maximum likelihood in each bin (black points). The error bars indicate an increase of the likelihood by $2\Delta\ln\mathcal{L} = 1$. 
The best-fit spectrum of the joint likelihood without an ALP  
(with an ALP with $m_\mathrm{neV} = 1.2$ and $g_{11} = 1$) 
is shown as a light (dark) red solid line.
}
\end{figure}
\fi
We simulate $N_B = 500$ random realizations of the turbulent field $\mathbf{B}_j$, $j=1,\ldots,N_B$.
The dependence of the likelihood on the realizations is not easily parametrizable and we cannot assume that the simulations map the space of possible realizations. 
Therefore, instead of profiling, we sort the $B$-field realizations for each tested $(m_a, g_{a\gamma})$ pair 
by increasing values of the 
product over the likelihoods $\mathcal{L}_i$
 and use the realization that corresponds to the $Q_B = 0.95$ quantile
of the likelihood distribution (profiling would correspond to $Q_B = 1$). 
We will denote this realization as $\mathbf{B}_{95}$ and the corresponding expected counts with $\boldsymbol{\mu}_{95}$.
Note that $\mathbf{B}_{95}$ might be different for different ALP parameters, so that $\mathbf{B}_{95}\equiv\mathbf{B}_{95}(m_a, g_{a\gamma})$.

Similar to \cite{wouters2013}, 
we evaluate the ALP hypothesis with a likelihood ratio test.
The test statistic ($\mathrm{TS}$) for the ALP hypothesis is calculated from the joint likelihood of all event types:
\ifdefined\fulltext
\begin{equation}
\mathrm{TS} = -2\sum\limits_i\ln\left(\frac{\mathcal{L}_i(\boldsymbol{\mu}_0, \hat{\hat{\boldsymbol{\theta}}} | \mathbf{D}) }
{\mathcal{L}_i(\hat{\boldsymbol{\mu}}_{95}, \hat{\boldsymbol{\theta}} | \mathbf{D}) } \right),
\end{equation}
\fi
where $\boldsymbol{\mu}_0$ are the expected counts for the null (no ALP) hypothesis with maximized nuisance parameters 
$\hat{\hat{\boldsymbol{\theta}}} \equiv {\hat{\boldsymbol{\theta}}}(\boldsymbol{\mu}_0)$
and $\hat{\boldsymbol{\mu}}_{95}$ are the expected counts under the ALP hypothesis
that, together with $\hat{\boldsymbol{\theta}}$, maximize  
the likelihoods of each event type.
We test ALP parameters on a logarithmic $(m_a,g_{a\gamma})$ grid with  $(19\times12)$ steps where
 $0.07 \leq m_{a,\mathrm{neV}} \leq 100$ and $0.1\leq g_{11} \leq 7$.
The mass range is chosen such that $E_\mathrm{crit}$ falls into the analyzed energy range 
whereas the maximum coupling is motivated by the bound found in \cite{ayala2014}.
 For the lower bound, the amplitude of the irregularities is too small to be detectable. 

In order to convert the $\mathrm{TS}$ value into a significance, we need to know the underlying probability distribution.
We derive the null distribution from Monte-Carlo simulations
and from it the threshold $\mathrm{TS}$ value, $\mathrm{TS}_\mathrm{thr}$, for which we
can reject the null hypothesis (see the Supplemental Material for details). 
For a rejection of the no-ALP hypothesis at a $3\,\sigma$ (global) significance level, we find
that $\mathrm{TS} > \mathrm{TS}_\mathrm{thr} = 33.1$.

\section{Results}
\label{sec:results}


The best-fit ALP parameters are found at $m_\mathrm{neV} = 44.6$ and $g_{11} = 4.76$  with 
 $\mathrm{TS} = 10.40 < \mathrm{TS}_\mathrm{thr}$,  and hence
the best fit with ALPs is not significantly preferred over the null hypothesis. 
We set upper limits by stepping over the ALP parameters
and calculating the difference $\lambda(m_a,g_{a\gamma})$ between the log-likelihood values for each pair $m_a,\,g_{a\gamma}$ and the best fit.
ALP parameters are excluded with $95\,\%$ confidence if $\lambda > \lambda_\mathrm{thr} = 22.8$.
The threshold value $ \lambda_\mathrm{thr}$ is calculated under the assumption
that the probability distribution of the alternative hypothesis follows the null distribution.
We have tested this assumption with simulations and found that this choice results in 
over coverage
for ALP parameters causing the strongest irregularities, thus yielding conservative limits.

\begin{figure*}[thb]
\ifdefined\fulltext
\includegraphics[width = .49\linewidth]{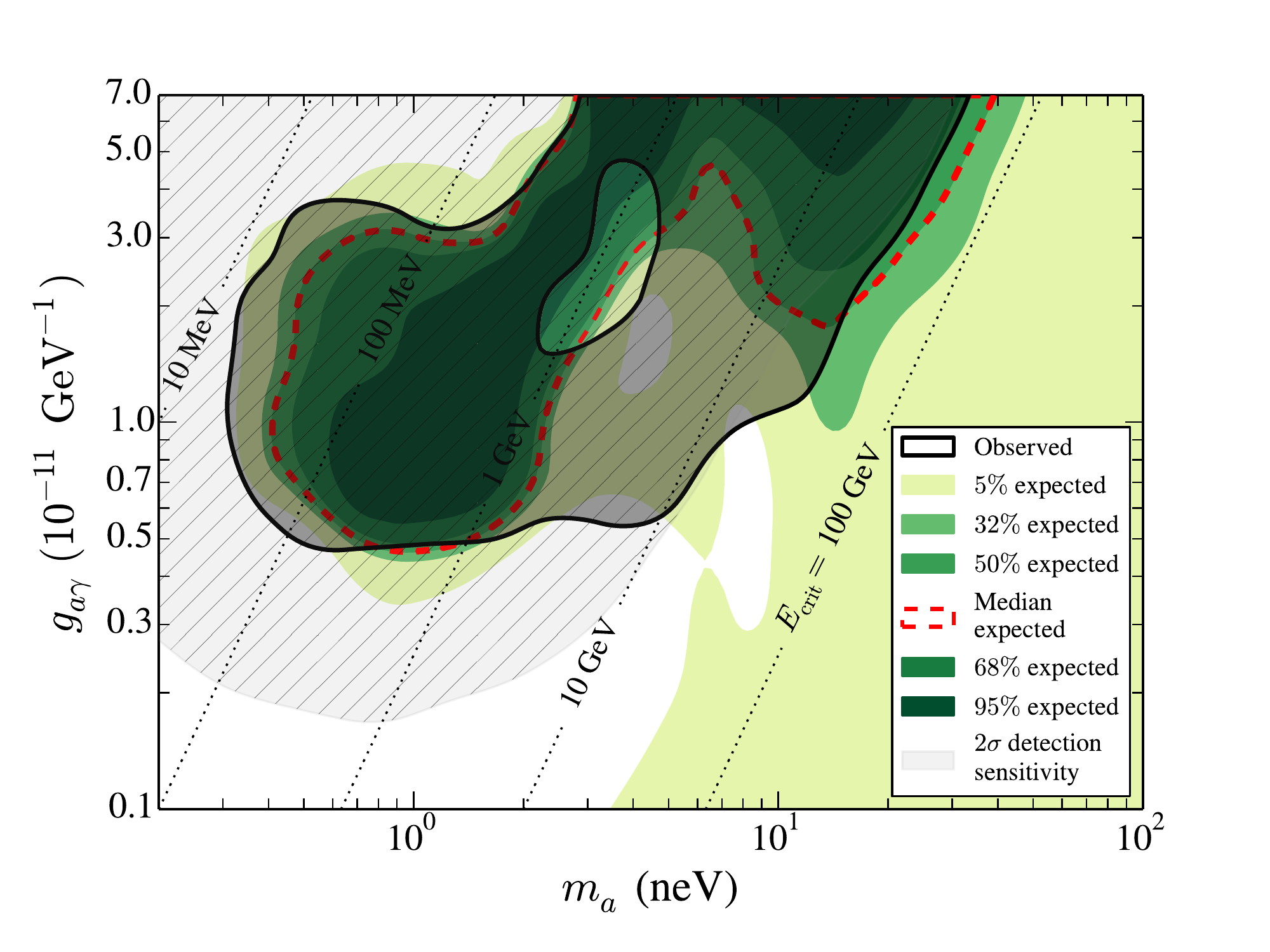}
\includegraphics[width = .49\linewidth]{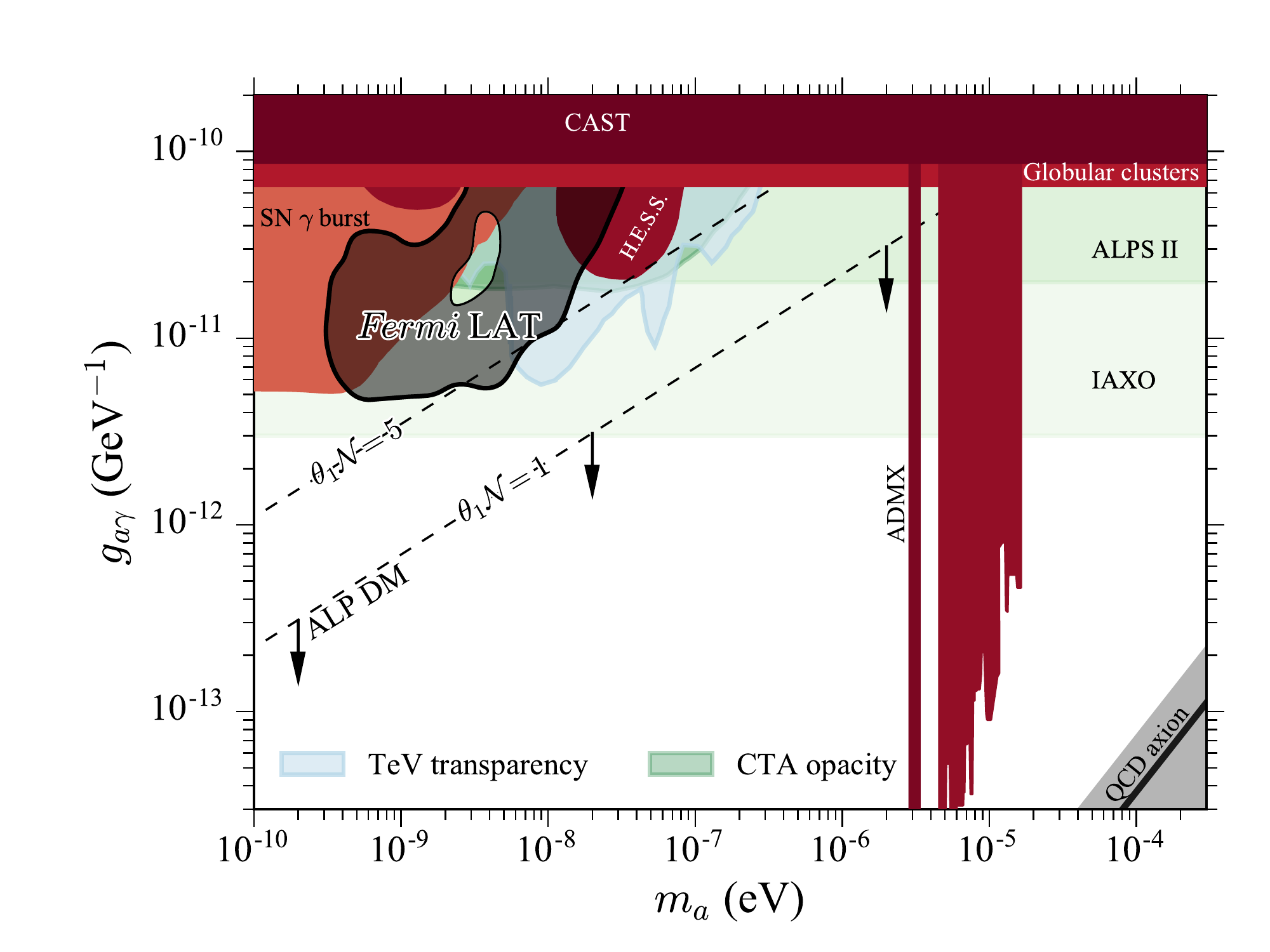}
\fi
\caption{\label{fig:results}\textit{Left}: Observed and expected 95\,\% confidence limits on the ALP parameters from 400
Monte-Carlo simulations.
Dotted lines correspond to constant critical energies.
The hatched gray region shows the parameters where ALPs are detectable at the $2\,\sigma$ confidence level (median sensitivity). 
\textit{Right}: Comparison of \emph{Fermi}-LAT limits with other works. 
Other Limits are shown in red, expected sensitivities in green.
The parameter space where ALPs could explain a low $\gamma$-ray opacity is shown in blue. 
ALPs below the $\mathcal{N}\theta_1 = 1$ line could account for all the DM. 
The QCD axion is shown as a gray shaded band and solid black line.   
See, e.g. \cite{meyer2014cta} and references therein.
}
\end{figure*}

The excluded parameter space is shown in the left panel of Fig. \ref{fig:results} (black shaded region). 
Photon-ALP couplings are ruled out between $0.5 \lesssim g_{11} \lesssim 3$ for $0.5 \lesssim m_{a,\mathrm{neV}} \lesssim 5$
and $g_{11} \gtrsim 1$ for $5 \lesssim m_{a,\mathrm{neV}} \lesssim 10$.
At high masses, the limits run almost parallel to the lines of constant $E_\mathrm{crit}$ 
(shown as dotted lines for $B_{\mu\mathrm{G}} = 10$).
For lower masses, 
 ALP couplings along the $E_\mathrm{crit} = 1\,$GeV line
with $1.3 \lesssim g_{11} \lesssim 4$ are not excluded.
Around this ``hole''-like feature, $P_{\gamma\gamma}$ exhibits rapid fluctuations for  
 almost the entire \emph{Fermi}-LAT energy range. 
Given the Poisson noise in the data, these ALP
parameters cannot be excluded.
We stress that the fit with
ALPs is not preferred over the null hypothesis.
For masses below $m_{a,\mathrm{neV}} = 0.5$, 
irregularities still enter the \emph{Fermi}-LAT energy range allowing to exclude ALP parameters.

The observed limits agree well with the expected exclusion region
derived from Monte-Carlo simulations  
(shaded regions).
The ``hole'' feature is not visible in the expected limits but occurs in certain Monte-Carlo realizations 
(an example is given in the Supplemental Material).  
In 5\,\% of the simulations (yellow shaded region), ALP parameters are excluded 
for which the $E_\mathrm{crit} > 100$\,GeV. 
This is expected since we have derived $\lambda_\mathrm{thr}$ from the null distribution where for 5\,\% 
of the simulations  one finds $\mathrm{TS} > \lambda_\mathrm{thr}$. 
The parameters for which we could detect an ALP signal at a  $2\,\sigma$ level
 agree well with the observed limits (gray hatched region;
 see the Supplemental Material for details).

The results are subject to systematic uncertainties related to the analysis and magnetic 
field parameters. 
Concerning the analysis,
changing the energy dispersion has the strongest effect on the limits. 
If we conservatively broaden the energy dispersion by 20\,\% 
the area of the tested ALP parameter grid with $\lambda > 22.8$ decreases by 25\,\%.
All other tested effects related to the analysis change the limits at most by $\sim4$\,\%.
Concerning the choice of $B$-field parameters, neither the  strength, the power spectrum, nor the 
dependence on the electron density of the magnetic field are well established for Perseus.
Therefore, the full analysis is repeated for a magnetic-field strength of $\sigma_B = 20\,\mu$G, 
for a Kolmogorov-type turbulence spectrum, $q = -11 / 3$ (as found in the cool-core cluster Hydra A, \cite[e.g.][]{kuchar2011}),
and by conservatively assuming that the magnetic field is zero beyond $r_\mathrm{max} = 100\,$kpc.
Increasing $\sigma_B$ increases the excluded area by 43\,\%. 
In comparison, the other tested parameters have a subdominant effect of maximally 16\,\%. 
The dependence of the limits on the particular choice of the EBL model 
is negligible due to the relative proximity of NGC\,1275 ($z = 0.017559$).
The absorption is maximally $\sim8\,\%$ at 500\,GeV 
with significantly smaller relative differences for a number of EBL models
\cite{dominguez2011,franceschini2008,finke2010,kneiske2010,gilmore2012,inoue2013}.
We provide a comprehensive summary of all tested systematic uncertainties in the Supplemental Material.

The limits derived in this work are compared to other limits and sensitivities of future experiments 
in Fig. \ref{fig:results} (right). Our results give the strongest constraints to date for $0.5 \lesssim m_{a,\mathrm{neV}} \lesssim 20$
and surpass the expected limits for the planned ALPS II  experiment \cite{alpsII} in that range.
They are only a factor of $\sim2$ below the exclusion prospects of the planned IAXO experiment \cite{irastorza2013}.
We note that the systematic uncertainties of the future experiments are likely to be smaller than the ones that 
apply to the present analysis. 
In conjunction with 
other limits taken at face value 
\cite{hess2013:alps,wouters2013,payez2015},
 the parameter space where ALPs could explain hints for a lower $\gamma$-ray opacity
  compared to EBL-model predictions (light blue region, \cite{meyer2013})
is now strongly constrained.
The limits do not constrain ALPs that could make up the entire DM content of the Universe. 
This corresponds to the region in Fig. \ref{fig:results} (right) below the $\theta_1\mathcal{N} = 1$ line,
where $\mathcal{N}$ is a model dependent factor 
and $\theta_1$ is the misalignment angle
\cite{arias2012}. 
Our analysis only constrains 
 ALPs that make up less than 4\,\% of the DM, or equivalently $\theta_1\mathcal{N} >5$.

Observations with future $\gamma$-ray instruments could improve the reported limits and test ALP DM models.
The planned Gamma-400 satellite, with an envisaged energy resolution of 1\,\% above 10\,GeV  \cite{cumani2015}, might be able to better resolve the spectra and probe higher ALP masses.
Higher masses could also be reached with the future Cherenkov Telescope Array (CTA) \cite{cta2011}.

It will be possible to reduce the uncertainties of the intra-cluster $B$ field with the upcoming Square Kilometer Array (SKA) that will conduct a full-sky polarisation survey \cite{gaensler2004}.
It is expected that SKA will observe
 hundreds of RMs of background sources for the most massive clusters,
thereby enabling a more precise determination of their magnetic fields \cite{bonafede2015}.

The analysis presented here can be easily extended to other sources that reside in clusters (e.g. M\,87 in the Virgo cluster) or in general to any source where ALP-induced spectral irregularities are expected. 
ALP parameters not constrained in the present analysis (such as those of the ``hole''-like feature)
could be probed with the different $B$-field configurations in other sources.

\ifdefined\fulltext
\begin{acknowledgments}
\section{Acknowledgments}
The \textit{Fermi}-LAT Collaboration acknowledges support for LAT development, operation and data analysis from NASA and DOE (United States), CEA/Irfu and IN2P3/CNRS (France), ASI and INFN (Italy), MEXT, KEK, and JAXA (Japan), and the K.A.~Wallenberg Foundation, the Swedish Research Council and the National Space Board (Sweden). Science analysis support in the operations phase from INAF (Italy) and CNES (France) is also gratefully acknowledged.
\end{acknowledgments}
\fi

\bibliographystyle{apsrev4-1}
%

\ifdefined\arxiv
\clearpage

\appendix

\setcounter{equation}{0}

\widetext
\begin{center}
  {\bf \large Supplemental Material:  Search for Spectral Irregularities due to Photon--Axionlike-Particle Oscillations with the Fermi Large Area Telescope}
\end{center}
\section{Energy dispersion and spectral binning}

Event reconstructions prior to {\pass} allowed to perform an independent analysis of $\gamma$-ray events 
that convert in either the \emph{front} or \emph{back} part of the LAT tracker, e.g. \cite[][]{ackermann:2012performance}.
With {\pass} this is now supplemented with additional \emph{event types} that make it possible 
to select events based on the quality of the angular reconstruction (PSF event types), or the quality of the energy reconstruction (EDISP event types). 
Event types within the PSF or EDISP partition are mutually exclusive and each energy energy bin contains about the same number 
of events of each type. 
The analysis presented here is performed for each EDISP type independently. 
The EDISP event types are defined through the cuts on energy dependent boosted decision tree variables used for the event classification
that determine the quality of the event reconstruction.

The exposure-averaged
 energy dispersion matrix for each event type EDISP$i$ ($i=0,1,2,3$) $\mathcal{D}^i_{kk'}$ 
 transforms the number of counts in true energy ($E$) $n_k$ in the $k$-th energy bin to the number of
counts in the $k'$-th bin of reconstructed energy ($E'$), $n_{k'}$, via
\begin{equation}
n^i_{k'} = \sum\limits_k \mathcal{D}^i_{kk'} n^i_k.
\end{equation} 
We extract $\mathcal{D}^i_{kk'}$ for six years of observation of NGC\,1275 
between 30\,MeV and 1\,TeV
 in 160 logarithmic-spaced energy bins
and evaluate it at the ROI center, i.e. the position of NGC\,1275.
The resulting matrices are shown in Fig. \ref{fig:edisp} (left). Clearly, the energy resolution improves from event type EDISP0 to EDISP3.

From the $\mathcal{D}^i_{kk'}$ we also compute the 68\,\% containment half-width 
energy resolution $\Delta E / E$ 
in the same way as described in \cite{ackermann:2012performance} (Fig.  \ref{fig:edisp}, right). 
At 10.2\,GeV the energy resolution improves from 16\,\% for EDISP0 to 4\,\% for EDISP3 (using the half width of $\Delta E / E$).
From the full-width energy resolution, 
we choose the number of logarithmic spaced energy bins used in the analysis. 
The number of bins should be small
 enough to resolve the photon-ALP spectral features after convolving them with the energy dispersion.
As a compromise between computational time and energy resolution, we
 choose a bin size of 30\,\% of the median energy resolution $\langle \Delta E / E\rangle$. 
The total number of bins for each event type EDISP$i$ is then 
\begin{equation}
N_\mathrm{bin} = \left\lfloor \frac{\log_{10}(E_\mathrm{max} / E_\mathrm{min})}{0.3 \langle \Delta E / E\rangle^i}\right\rfloor,
\end{equation}
with $E_\mathrm{min} = 100\,$MeV and $E_\mathrm{max} = 500\,$GeV.
Since we forward fold the model spectrum with the detector response (Eq. (2) in the main 
article), bin-by-bin correlations are self-consistently accounted for.

\begin{figure}[thb]
\includegraphics[width = .49\linewidth]{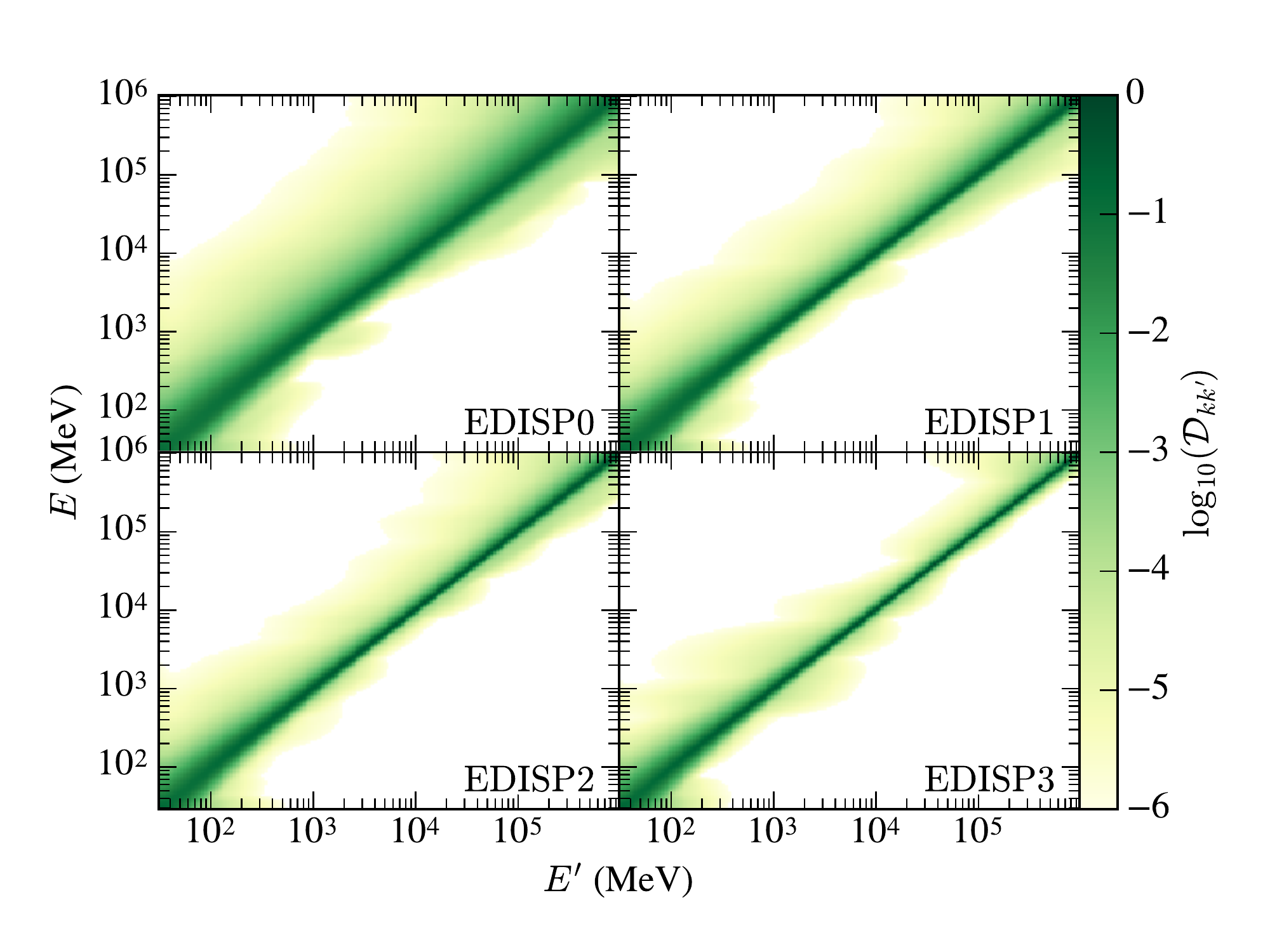}
\includegraphics[width = .49\linewidth]{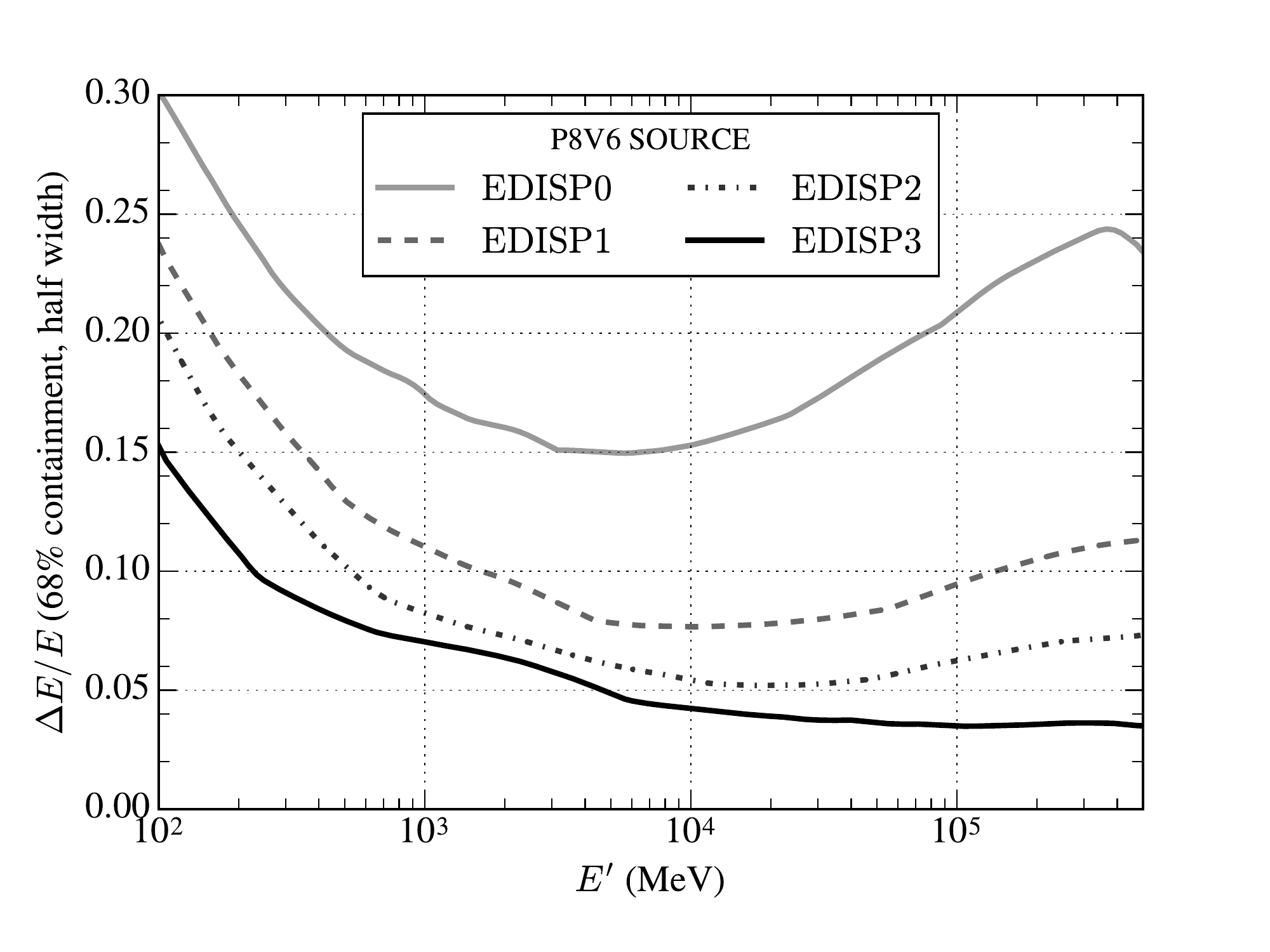}
\caption{\label{fig:edisp}\textit{Left:} 
Exposure-weighted
 energy dispersion matrices for 
 the four EDISP types in {\pass}.
  \textit{Right}: Energy resolution (half-width of the 68\,\% containment) of all  event types 
derived from the energy dispersion matrices.}
\end{figure}

\section{Modelling Photon-ALP oscillations}
We describe the photon-ALP beam of energy $E$ in terms of the density matrix $\rho = \Psi \otimes \Psi^\dagger$, 
where $\Psi = (A_1,A_2,a)^T$ is a photon-ALP state with photon polarizations $A_{1,2}$ and the ALP field strength $a$.
Assuming the beam to propagate along the $x_3$ direction and neglecting photon absorption, 
the evolution of the system along $x_3$ is described with the Von-Neumann-like commutator equation, e.g. \cite[][]{mirizzi2009},
\begin{equation}
i\frac{\mathrm{d\rho}}{\mathrm{d}x_3} = \left[\rho,\mathcal{M}\right], \label{eq:neumann}
\end{equation}
with the mixing matrix $\mathcal{M}$.
The photon-ALP mixing is induced by the transversal magnetic field $\mathbf{B}_\perp$ only. 
For a homogenous field orientated along $x_2$, $\mathbf{B}_\perp = B \mathbf{\hat{e}}_2$, the mixing matrix reads \cite{raffelt1988,mirizzi2007},
\begin{equation}
\mathcal{M} = 
\begin{pmatrix}
\Delta_{\perp} & 0 & 0\\
0 & \Delta_{||} & \Delta_{a\gamma} \\
0 & \Delta_{a\gamma} & \Delta_a
\end{pmatrix},
\end{equation}
where Faraday rotation is neglected. The $\Delta$ terms give the momentum differences between the polarization states.
The terms $\Delta_\perp = \Delta_\mathrm{pl} + 2\Delta_\mathrm{QED}$ and $\Delta_{||} = \Delta_\mathrm{pl} + (7/2)\Delta_\mathrm{QED}$
describe the propagation of photons in a plasma with plasma frequency $\omega_\mathrm{pl}$, where $\Delta_\mathrm{pl} = -\omega_\mathrm{pl}^2 / 2E$.
They also include the QED vacuum polarization through $\Delta_\mathrm{QED}  = \alpha E B^2 / 45 \pi B^2_\mathrm{cr}$, with 
the critical magnetic field $B_\mathrm{cr} \sim 4.4\times10^{13}\,$G and the fine-structure constant $\alpha$.
For the energies and magnetic fields under consideration, this term is, however, completely negligible. 
The same holds for an additional photon dispersion term \cite{dobrynina2015} since we neglect photon-ALP mixing in the intergalactic medium beyond the galaxy cluster. 
 Photon-ALP mixing is induced by the off-diagonal term 
$\Delta_{a\gamma} = g_{a\gamma} B / 2$. The kinetic term for the ALP is $\Delta_a = -m_a^2 / 2E$. 
Numerical values for the momentum differences in suitable units are given in e.g. \cite{horns2012ICM}.
The oscillation length is given by $L = 2\pi / \Delta_\mathrm{osc}$, where $\Delta_\mathrm{osc} = [(\Delta_a - \Delta_{||})^2 - 4\Delta_{a\gamma}^2]^{1/2}$.
For a 10\,GeV $\gamma$ ray and $B_{\mu\mathrm{G}} = 10$, $m_{a,\mathrm{neV}} = 10$, and $g_{11} = 1$, one finds $L \sim 7.5$\,kpc.
For the considered profile of the electron density (Eq. (4) in \cite{churazov2003}),
 the plasma frequency is at most 10\,\%
 of the minimum ALP mass considered. 
 Therefore, $\Delta_a - \Delta_{||} \sim \Delta_a$, 
 making the critical energy 
 (Eq. (1) in the main article) essentially independent of $\omega_\mathrm{pl}$.

Equation \eqref{eq:neumann} is solved by means of the transfer matrix $\mathcal{T}$, 
$\rho(x_3) = \mathcal{T}(x_3,0; E)\rho(0)\mathcal{T}^\dagger(x_3,0; E)$, with the initial condition 
$\mathcal{T}(0,0;E) = 1$, e.g. \cite[][]{csaki2003,deangelis2007,sanchezconde2009,bassan2010}.
Neither the magnetic field in the Perseus cluster nor the Galactic magnetic field are homogeneous. 
Therefore, the path in the different $B$ fields is split up into $N$ segments. In each segment, 
$\mathbf{B}_\perp$ is assumed to be constant and forms an angle $\psi_i$, $(i = 1,\ldots,N)$ with the $x_2$ axis. The $B$ field  is modeled as a a divergence-free homogeneous isotropic field with Gaussian turbulence \cite{meyer2014}
The full transfer matrix is then
 \begin{equation}
\mathcal{T}(x_{3,N},x_{3,1};\psi_{N},\ldots,\psi_1;E) = \prod\limits_{i = 1}^{N} \mathcal{T}(x_{3,i+1},x_{3,i};\psi_{i};E).
\end{equation}
In this setup, the probability to observe a photon of either polarization $\rho_{jj} = \mathrm{diag}(\delta_{1j}, \delta_{2j},0)$ after the $N$-th domain is given by
\begin{equation}
P_{\gamma\gamma} = \sum\limits_{j = 1,2} \mathrm{Tr}\left(\rho_{jj} \mathcal{T} \rho_0 \mathcal{T}^\dagger\right).
\end{equation}
For the initial polarization, we assume an un-polarized pure $\gamma$-ray beam, $\rho_0 = (1/2) \mathrm{diag}(1,1,0)$. For the fiducial parameters of the Perseus cluster, 
the Galactic magnetic field of Ref. \cite{jansson2012}, and the EBL model of Ref. \cite{dominguez2011}, the photon survival probability for one ALP mass and 
coupling is shown 
in Fig. \ref{fig:pgg} for 500 random realizations of the cluster magnetic field. 

\begin{figure}
\includegraphics[width = .9 \linewidth]{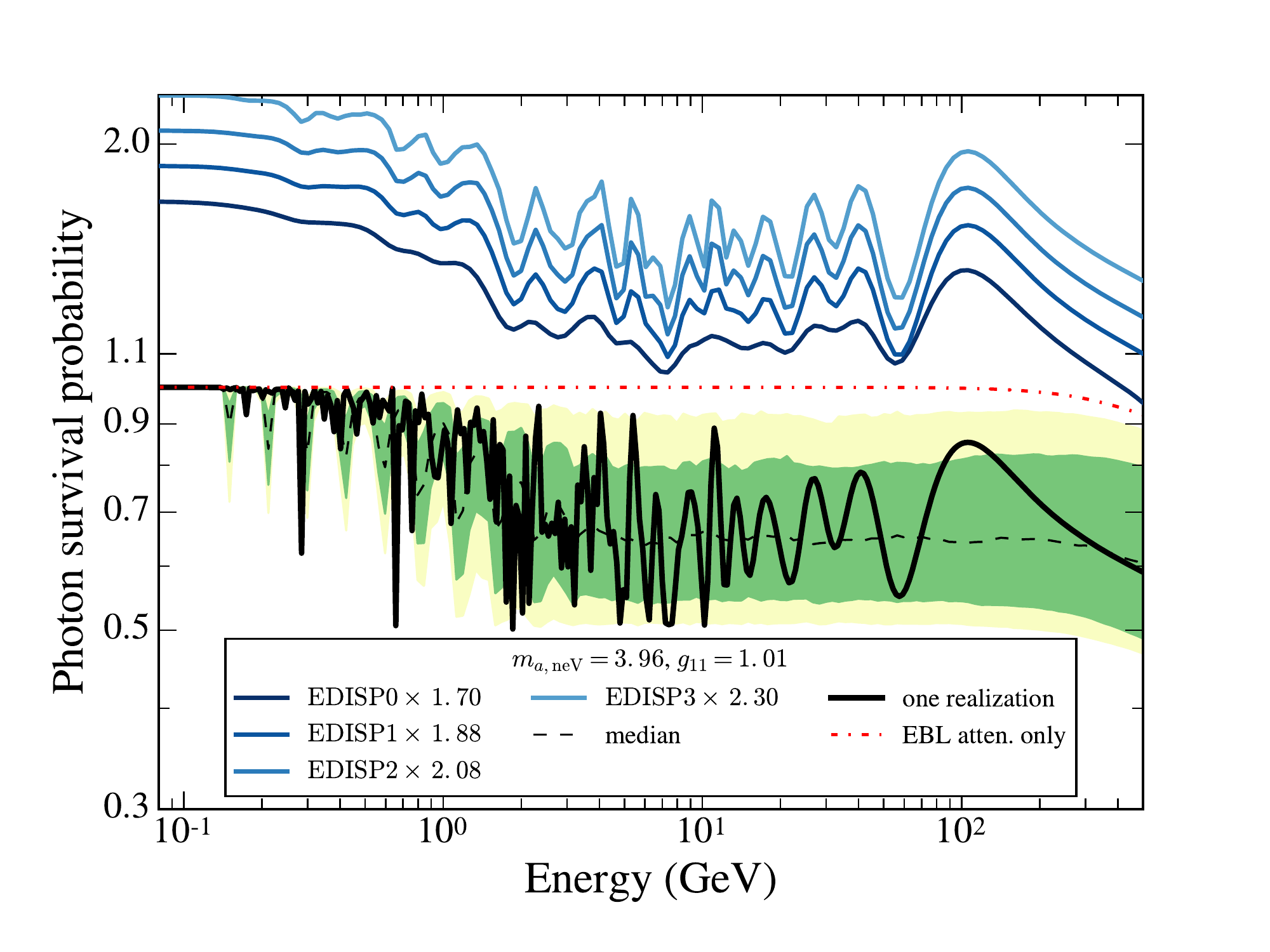}
\caption{\label{fig:pgg}Example of the photon survival probability for an ALP with mass $m_{a,\mathrm{neV}} = 3.96$ and coupling $g_{11}=1.01$. 
The green (yellow) shaded region shows the 68\,\% (95\,\%) 
containment of 500 realizations of the turbulent cluster $B$ field. The blue lines show $P_{\gamma\gamma}$ for one realization (black solid line)
convolved with the energy dispersion of the different event types (from bottom to top EDISP0-3). They are scaled upwards for better visibility.
The EBL attenuation sets in above 100\,GeV (red dash-dotted line).
}
\end{figure}

\section{Null distribution and confidence intervals}
The usual procedure to convert a $\mathrm{TS}$ value into a significance with which one can 
reject the null hypothesis using Wilks' Theorem \cite{wilks1938} 
cannot be applied in the present analysis. 
First of all, the spectral irregularities do not scale linearly with the ALP parameters. 
Secondly, under the null hypothesis, the likelihood values are independent of the magnetic-field realizations which are additional nuisance parameters under the alternative hypothesis. 
Thirdly, photon-ALP oscillations are completely degenerate in coupling and magnetic field since the oscillation strength scales with 
$\Delta_{a\gamma}$,
leading to potential correlations between the different tested ALP parameters. 
For these reasons, the null distribution is a priori unknown and needs to be derived from Monte-Carlo simulations. 

We simulate $N_\mathrm{PE} = 400\upnum$ six year data sets (pseudo experiments, PEs) 
within a $20^\circ\times20^\circ$ ROI including all 3FGL point sources and the isotropic backgrounds for each event type between 
30\,MeV and 1\,TeV. We perform these simulations with an energy window and ROI larger 
than the ones used in the analysis in order to account for a possible migration of photons between adjacent energy bins or sky regions due to the PSF and energy dispersion.
All point sources, including NGC\,1275, are simulated with energy dispersion, whereas we disable it for the diffuse templates.
The actual pointing history of the LAT is used in the simulations. No ALP signal is included.
We perform the exact same analysis as done on the data on each PE and calculate 
the $\mathrm{TS}$ values as the difference between the maximum likelihood and the profile likelihood 
of the fit without ALPs. 
For one PE, we show the resulting exclusion region in comparison with the data in Fig. \ref{fig:null-dist} (left).
The figure illustrates that the complex behavior of the exclusion region is already evident in Monte-Carlo simulations.
The full null distribution is shown in Fig. \ref{fig:null-dist} (right). 
It can be parametrized with a non-central $\chi^2$ distribution with 10.09 degrees of freedom (d.o.f.)
and non-centrality parameter $s = 2.51$. For these parameters the
 Kolmogorov-Smirnov test yields a $p$-value of 0.70. 
All $\mathrm{TS}$ values are larger than zero indicating that all fits including ALPs are preferred. 
This is expected since we have deliberately chosen a high quantile $Q_B = 0.95$. 
Selecting a smaller quantile would shift the distribution towards zero as $B$-field realizations would be selected 
that lead to a worse fit to the data. 
In order to avoid false ALP detections we use the null distribution to derive 
the threshold $\mathrm{TS}$ value, $\mathrm{TS}_\mathrm{thr}$, for which evidence for ALPs could be claimed at a 3\,$\sigma$ (2$\,\sigma$) confidence level.
From the fit to the null distribution
one finds that  
$\mathrm{TS}_\mathrm{thr} = 33.1^{+ 5.3}_{-5.8}\upnum$ ($23.1^{+1.8}_{-1.2}$).
The uncertainties are at 68\,\% confidence and are derived from bootstrapping the null distribution $10^4$ times.  

\begin{figure}[thb]
\includegraphics[width = .49\linewidth]{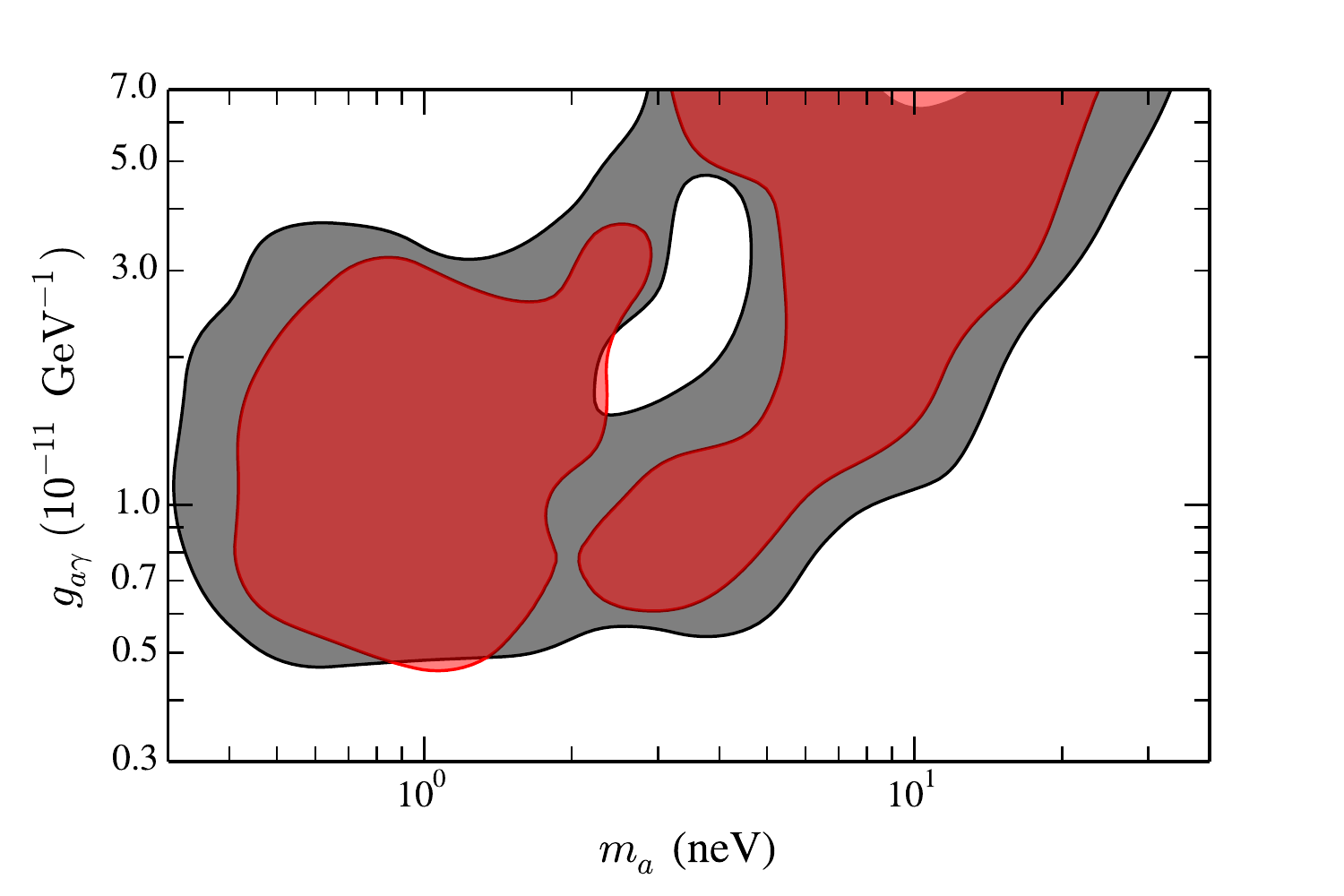}
\includegraphics[width = .49\linewidth]{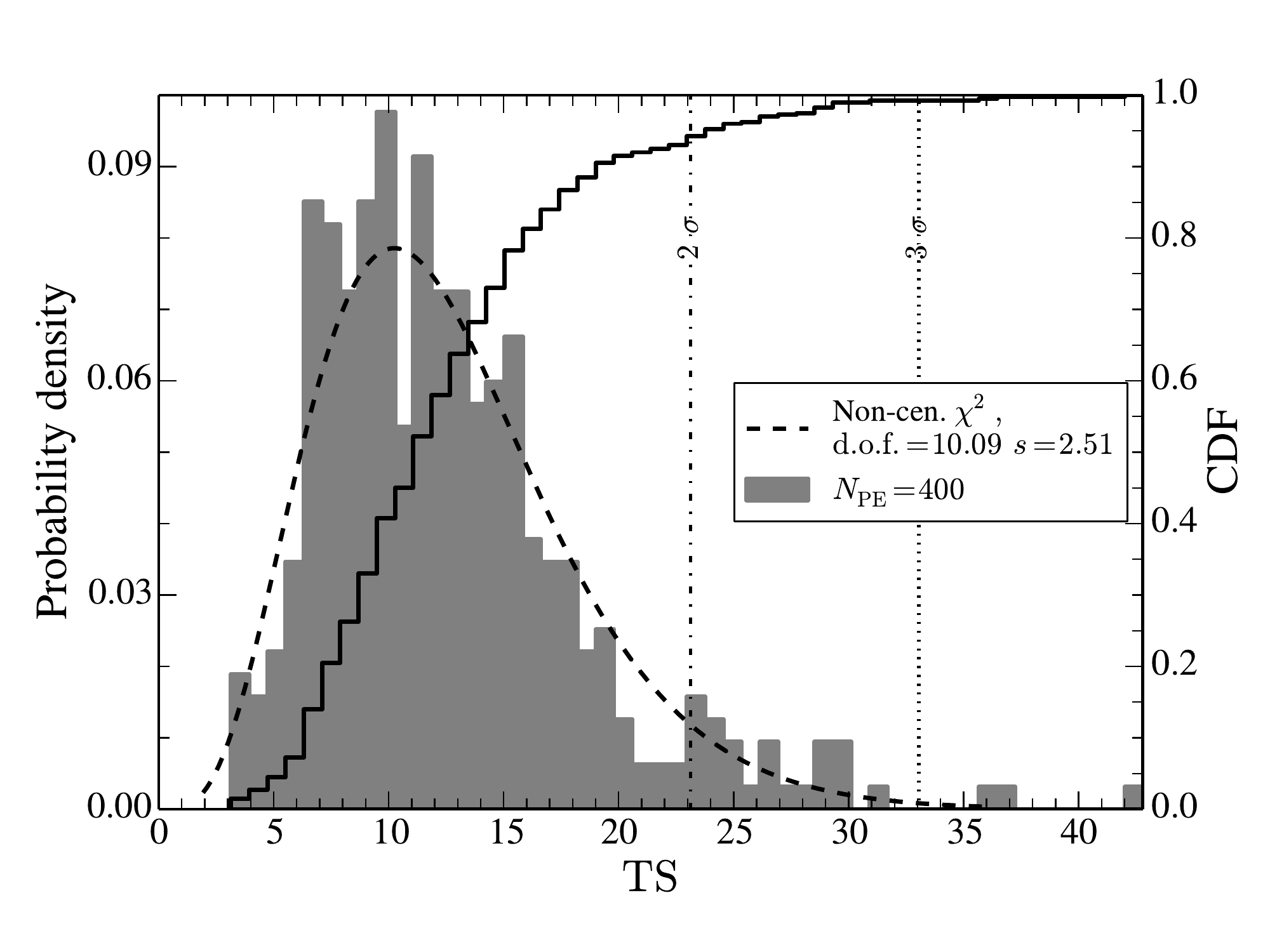}
\caption{\label{fig:null-dist} \textit{Left}: 
Excluded parameter space from the data (black shaded region), 
and for one PE (red shaded region).
\textit{Right}: Null distribution of $\mathrm{TS}$ values
 for $N_\mathrm{PE} = 400$ and $Q_B = 0.95$. The dashed line represents 
a fit to the data with a non-central $\chi^2$ distribution. 
The cumulative distribution function (CDF) is shown as a black solid line.
The dash-dotted (dotted) vertical line shows the threshold $\mathrm{TS}$ value that 
corresponds to a $2\,\sigma$ ($3\,\sigma$) confidence level.
}
\end{figure}

Since we are deriving  $\mathrm{TS}_\mathrm{thr}$ from simulations, one could ask why we are choosing $Q_B = 0.95$ 
instead of profiling, i.e., $Q_B = 1$.
The reason is that we can only test a limited number of $B$-field realizations, thereby probably missing the one actually realized in nature. 
While the profile likelihood ($Q_B = 1$) is sensitive to the number of tested $B$-field realizations, 
we find that with  $Q_B = 0.95$ our analysis is insensitive to the ad-hoc choice of number of realizations. 
Furthermore, we find better coverage for this choice, as discussed below.

We derive the ALP parameters that are detectable at a certain significance 
using an Asimov data set, i.e. setting the observed counts equal to the  
expected number of counts \cite{cowan2011}.
For each ALP parameter and $B$-field realization we calculate the expected number of counts using Eq. (2) of the main article
assuming the spectral parameters of NGC\,1275 that are given in the 3FGL. 
The obtained Asimov data are subsequently fitted with a log parabola without an ALP contribution.
As before, for each tested ALP parameter pair, we sort the likelihood values for the different $B$-field realizations and evaluate the test statistic for the realization that corresponds to the $Q_B = 0.95$ quantile,
\begin{equation}
\mathrm{TS}_\mathrm{A} = -2\sum\limits_i\ln\left(\frac{\mathcal{L}_i(\boldsymbol{\mu}_0, \hat{\hat{\boldsymbol{\theta}}} | \boldsymbol{\mu}_{95}) }
{\mathcal{L}_i(\hat{\boldsymbol{\mu}}_{95}, \hat{\boldsymbol{\theta}} |\boldsymbol{\mu}_{95}) } \right),
\end{equation}
where the subscript A denotes the Asimov data set. 
An ALP signal could be detectable at a $2\,\sigma$ level where $\mathrm{TS}_\mathrm{A} > 23.1$, shown as the gray hatched region in Fig. 2 (left) of the main article.

For the same reasons that make it necessary to derive the null distribution from Monte-Carlo simulations, 
we cannot assume that it is possible to derive upper limits on the ALP mass and coupling by increasing 
the log likelihood by $\lambda_\mathrm{thr} = 5.99$.\footnote{This value 
corresponds to a $2\,\sigma$ confidence limit in case of a $\chi^2$ distribution with 2 d.o.f. (ALP mass and coupling).}
The $\lambda$ values are given by the log-likelihood ratio test between the likelihood with a certain ALP mass and coupling  with 
conditionally maximized nuisance parameters ${\hat{\boldsymbol{\theta}}}(m_a,g_{a\gamma}, \mathbf{B}_{95})$ and the unconditionally maximized likelihood, 
\begin{equation}
\lambda(m_a,g_{a\gamma}) = -2\sum\limits_{i,k'}\ln\left(\frac{\mathcal{L}(m_a,g_{a\gamma}, \mathbf{B}_{95}, {\hat{\boldsymbol{\theta}}}(m_a,g_{a\gamma}, \mathbf{B}_{95})  | D_{ik'})}
{\mathcal{L}(\hat{m}_a,\hat{g}_{a\gamma}, \mathbf{B}_{95}, \hat{\boldsymbol{\theta}} | D_{ik'})} \right).
\end{equation}
In the above equation, $\hat{m}_a$, $\hat{g}_{a\gamma}$, and $\hat{\boldsymbol{\theta}}$ denote the maximum likelihood estimators. 
As before, we have chosen the magnetic field $\mathbf{B}_{95}$ that corresponds to the $Q_B = 0.95$ quantile 
of the distribution of likelihoods with respect to the $B$-field realizations.
Ensuring coverage would  in principle require to simulate the distribution of $\lambda(m_a,g_{a\gamma})$ (\textit{alternative distribution}) for the complete parameter space, which is not feasible.
We therefore make the ansatz that the alternative distribution for each tested ALP parameter is equal to the null distribution.
From the non-central $\chi^2$ fit to the null distribution we find that 
 $\lambda_\mathrm{thr} = 22.8\upnum$ corresponds to a 95\,\% upper limit (cf. Fig. \ref{fig:null-dist}, right).
For this choice, the expected limits (median sensitivity) agree well 
with the ALP parameters for which a signal could be detected (see Fig. 2, left, of 
the main article). 
We examine the coverage with simulations where we multiply the intrinsic spectrum of NGC\,1275 with the photon survival probability including ALPs.
We test seven 
ALP parameter pairs with masses between 1 and 60\,neV and couplings between 0.5 and $7\times10^{-11}\,\mathrm{GeV}^{-1}$,
and simulate 100 PEs for each parameter pair.
By construction, we find correct coverage for ALP parameters that result in weak irregularities or irregularities largely outside 
the \emph{Fermi}-LAT energy range. In these cases we expect the alternative distribution to be close to the null distribution.  
For masses and couplings where the ALP effect is strongest, we consistently find over coverage, thus making our results conservative.
The over coverage becomes more severe if we select $Q_B = 0.5$ or if we treat the magnetic field in a Bayesian fashion by taking the average over the $B$-field realizations. 
For $Q_B = 1$, the trend reverses and we find under coverage. 

We have also carried out a Bayesian analysis with flat priors on the magnetic-field realizations
and logarithmic flat priors for $m_a$ and $g_{a\gamma}$. 
We find the limits corresponding to a $(1 - \alpha)$ confidence level
by integrating over the posterior probability where we sort by decreasing likelihood values.
Studying the coverage of the limits, we find under coverage for the same injected ALP signals as above. 
Therefore, we compare the 99.9\,\% C.L. Bayesian limits (resulting in coverage $>95\,\%$) with the frequentist limits which 
give coverage $>95\,\%$ in Fig. \ref{fig:bayes}.
The Bayesian limits exclude 3\,\% less of the parameter space than the frequentist analysis. The shape
of the excluded region is comparable in the two analyses.

\begin{figure}
\includegraphics[width = 0.7\linewidth]{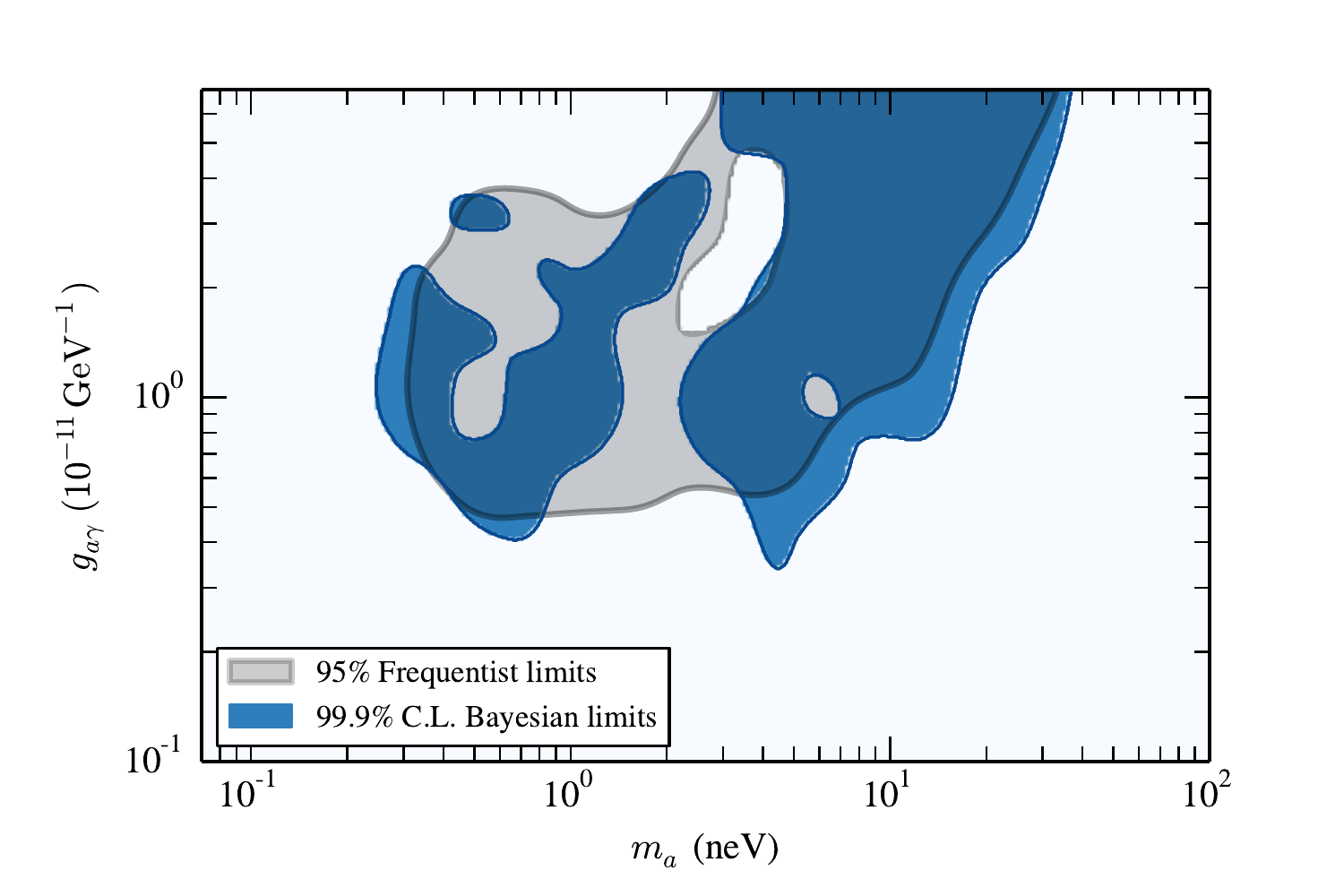}
\caption{\label{fig:bayes}Comparison between 99.9\,\% Bayesian and frequentist upper limits. See text for further details.}
\end{figure}

\section{Systematic Uncertainties}
We investigate how the systematic uncertainties connected to the LAT instrument, to the analysis choices, 
and to the choices for the magnetic-field parameters affect the derived bounds on the ALP parameters. 
We quantify the importance of each systematic uncertainty by studying the change of the excluded area $\alpha _\mathrm{sys}$, i.e. the 
area of the tested ALP parameter space for which $\lambda > \lambda_\mathrm{thr}$.
All examined systematics and the respective values of $\alpha _\mathrm{sys}$ are summarized in  Tab. \ref{tab:sys_unc}. 
In the following, we discuss each source of systematic uncertainty in further detail. 

\begin{table}[thb]
\begin{tabular}{l|c}
\hline
\hline
Systematic test & $\alpha_\mathrm{sys}$  \\
\hline
Bin-by-bin power-law index & \upnum$-4$\,\% \\
Bracketing of the effective area & \upnum$-2$\,\%  \\
Broadening of energy dispersion by 20\,\% & \upnum$-25$\,\%  \\
\hline
Central magnetic field $\sigma_B = 20\,\mu$G &\upnum$+43$\,\% \\
Kolmogorov turbulence spectrum $q = -11 / 3$ & \upnum$+12$\,\%\\
Maximum spatial extent of magnetic field $r_\mathrm{max} = 100\,$kpc & \upnum$-16$\,\% \\
\hline
\end{tabular}
\caption{\label{tab:sys_unc} Summary of systematic effects and their effect on the ALP bounds
quantified by the change in the excluded area, $\alpha_\mathrm{sys}$,  of the ALP parameter space induced by the corresponding effect.
}
\end{table}

\subsubsection{LAT instrument and analysis}

\paragraph{Assumed power-law index for bin-by-bin likelihood.}
In the extraction of the likelihood curves as a function of the expected counts in each energy bin
we have assumed a power law with spectral index $\Gamma = 2$ for NGC\,1275.
 Since the bin size is chosen as 30\,\% of the median energy resolution, the bin-by-bin likelihood should essentially be independent 
of the assumed spectral shape. We test this by repeating the ALP the analysis for power-law indices of $\Gamma = 2.5$ and $\Gamma = 1.5$.
As expected, the excluded area in the tested parameter space only changes marginally.
Indeed, it decreases by $\alpha_\mathrm{sys} = \upnum4\,\%$ for $\Gamma = 1.5$ whereas it is unchanged for $\Gamma = 2.5$.

\paragraph{Bracketing of the effective area.}
 The systematic uncertainty for the effective area is assessed 
using the IRF bracketing method \cite{ackermann:2012performance}. 
The exposure is scaled with $(1 + \epsilon(E)B(E))$, where
$\epsilon(E)$ is the relative uncertainty of the effective area and we take $B(E) = \pm 1$. 
For an analysis making use of the individual event types it is recommended
to assume a relative uncertainty of 10\,\% between 100\,MeV and 100\,GeV. 
It increases linearly with the logarithm of energy to 20\,\% at 30\,MeV and 1\,TeV\footnote{\url{http://fermi.gsfc.nasa.gov/ssc/data/analysis/LAT_caveats.html}}.
Using this scaling and repeating the analysis, we find a decrease by
$\alpha_\mathrm{sys} = \upnum2\,\%$ for $B = +1$. 
For $B = -1$, the results remain unchanged.
The small change of the limits is anticipated since we are interested 
in spectral irregularities that are not affected by an overall scaling of the effective area.

\paragraph{Small-scale structures in the effective area.}
We assess the possibility that unmodeled structures exist in the effective area.
Such structures might mimic spectral irregularities induced by ALPs. 
We follow a similar approach as in \cite{fermi-lat:2015p8lines} and estimate 
the strength of such structures from the residuals of a spectral analysis of  bright $\gamma$-ray sources.
At low energies (100\,MeV--5\,GeV) we consider the Vela pulsar and for higher energies (5\,GeV--100\,GeV) the Earth limb. 
The basic data selections are summarized in Tab. \ref{tab:vela-limb}.\footnote{For the actual data analysis of the Vela pulsar and the Limb we use slightly broader energy windows.}
All other analysis choices are the same as for the main analysis of NGC\,1275.

For the Vela pulsar, we calculate the pulsar phases $\varphi$ using the standard ephemeris\footnote{\url{http://www.slac.stanford.edu/~kerrm/fermi_pulsar_timing/}} and the \emph{tempo2}  package\footnote{\url{http://www.atnf.csiro.au/research/pulsar/tempo2/}}.
The spectral parameters of the background sources are determined separately for each EDISP event type
in the off-pulse interval $\varphi \in [0.7,1]$ using the standard binned analysis pipeline of the \emph{Fermi}-LAT \emph{Science Tools}.
The parameters are taken as an input for the phase averaged fit including the Vela pulsar. 
As in the 3FGL, the Vela spectrum is modeled with a power law with a super-exponential cutoff.\footnote{
The spectral model for a power law spectrum with super-exponential cutoff is defined though
$\mathrm{d}N / \mathrm{d} E = N (E/E_0)^{-\gamma_1}\exp((E / E_c)^{\gamma_2})$.
}
We show the fit residuals over the entire ROI in Fig. \ref{fig:wiggles}.

The Earth limb $\gamma$ rays are produced by cosmic-ray interactions 
in the Earth's upper atmosphere.
Therefore, the limb data set predominantly comprises $\gamma$ rays that arrive at a rocking angle $> 52^\circ$
and in the zenith angle range $111.1002^\circ < \theta_z < 112.9545^\circ$ (\textit{On} region). 
For the astrophysical background, photons within $108.6629^\circ < \theta_z < 109.5725^\circ$ and $114.5193^\circ < \theta_z < 115.4675^\circ$ are considered (\textit{Off} regions). 
The zenith angle cuts are chosen such that the solid angle for the \textit{On} and \textit{Off} regions is equal to $0.06\pi$, respectively \cite{ackermann:2012performance}.
The expected number of counts in the energy bin of width $\Delta E_k$ for each event type $i$ is given by
\begin{equation}
\mu_{\mathrm{limb},{ik}} =  N_\mathrm{limb} \int\limits_{\Delta E_{k}} \mathrm{d}E (E / E_0)^{-\Gamma_1} \left( 1 + (E / E_b)^{(\Gamma_1 - \Gamma_2) / \beta }\right)^{-\beta}  \mathcal{E}^i(E),
\end{equation}
where $\mathcal{E}^i(E)$ is the exposure averaged over the full sky.
The limb spectrum is modeled with a smooth broken power law to account for 
the effect of the Earth's geomagnetic cutoff. 
Following \cite{albert2014}, we fix the parameters controlling the spectrum
below and around the cutoff, namely $E_b = 370.3\,$GeV, $\Gamma_1 = 1.5232$, and $\beta = 0.7276$,
whereas the normalization $N_\mathrm{limb}$ and the spectral index above the cutoff $\Gamma_2$ are free parameters. 
The reference energy is fixed to the central energy of the energy bin
where the number of excess counts ($n_\mathrm{excess} = n_\mathrm{On} - n_\mathrm{Off}$) divided by its uncertainty
is maximal. 
The expected number of counts of the background is calculated in the same way, assuming however 
a simple power law as the spectral model.
We neglect the energy dispersion since both spectra for the limb and the background 
are smooth functions in energy that do not exhibit a cutoff
in the considered energy range. 
The fit residuals above 5\,GeV are shown in Fig. \ref{fig:wiggles}.

We use the residuals to estimate the maximum possible effect of small-scale structures 
in the exposure in the following way. 
We define a correction function $c_i(E)$ for each event type $i$ that
is equal to $P_{\gamma\gamma}$ if the amplitude of $P_{\gamma\gamma}(E)$ is smaller than the absolute value of the 
fractional residual at energy $E$. Otherwise, 
$c_i(E)$ is equal to the maximum amplitude of the absolute value of the fractional residual at energy $E$. 
We then divide each exposure by the corresponding $c_i(E)$. 
In this way, irregularities with an amplitude smaller than the fractional residuals will be removed and 
irregularities with larger amplitudes will be reduced. 
The maximum amplitude of the $c_i$ functions is given by the maximum of the absolute value of 
the residuals in Fig. \ref{fig:wiggles}.
However, we find that the limits are practically unchanged. 
The reason is that ALP parameters that are excluded at $95$\,\% confidence  
lead to irregularities of a larger amplitude than what can be counter-acted by possible small scale structures in the exposure.

\begin{table}[thb]
\begin{tabular}{l|cc}
\hline
\hline
{} &Vela pulsar& Earth limb \\
\hline
Time range (MET) &239846401--403747203  & 239557417--428859819 \\
Energy range & 100\,MeV--10\,GeV & 1\,GeV--100\,GeV\\  
LAT rocking angle & -- &  $> 52^\circ$ \\ 
ROI size & $10^\circ\times10^\circ$ & Full sky \\ 
\hline
\end{tabular}
\caption{\label{tab:vela-limb}Basic data selection cuts for the analysis of the Vela pulsar and the Earth limb.}
\end{table}

\begin{figure}[htb]
\includegraphics[width = .85 \linewidth]{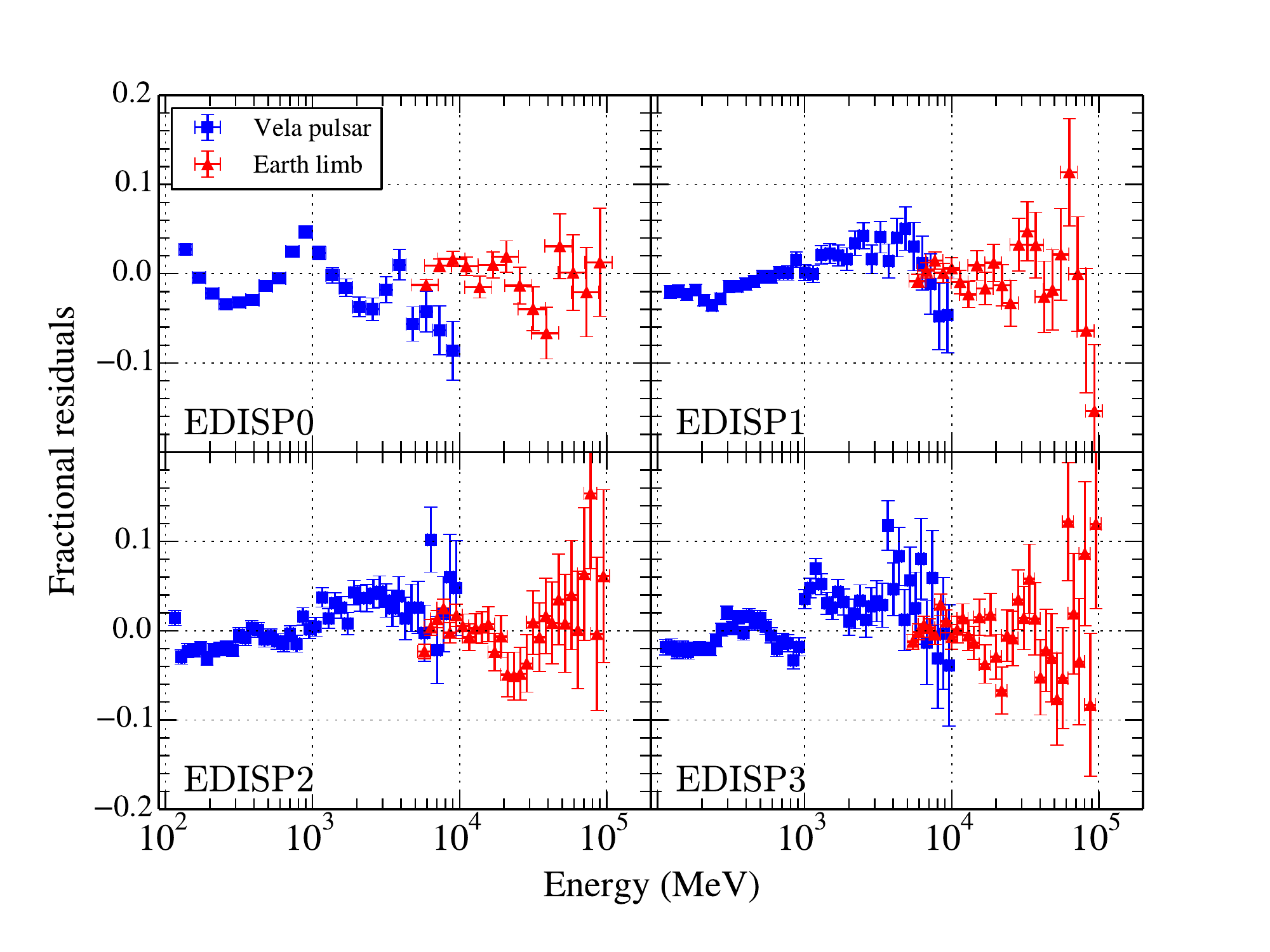}
\caption{\label{fig:wiggles}
Fractional residuals of the fits to the Vela pulsar and Earth limb used to estimate
the amplitude of spectral features induced by energy-dependent biases
  in the effective area.
The residuals are defined as the differences between the observed and expected counts divided by the expected counts.
}
\end{figure}

\paragraph{Broadening of the energy dispersion.}
We investigate the effect on the ALP limits in the case that energy resolution is worse than anticipated. 
The energy dispersion matrix of each event type is broadened along the reconstructed 
energy axis 
by the transformation $\mathcal{D}^i_{kk'}(E,E') \to (1 -\epsilon) \mathcal{D}_{kk'}^i(E,E' - \epsilon (E' - E))$.
The ALP analysis is repeated with the degraded energy dispersion for $\epsilon = 5\,\%,10\,\%$, and $20\,\%$.
The results are shown in Fig. \ref{fig:sys} (left panel). 
The fraction of excluded ALP parameters on the tested grid decreases
by 6\,\%, 11\,\%, and 25\,\%, respectively, for the choices of $\epsilon$ above.
The effect of the degraded energy dispersion is most visible around the ``hole''-like feature
where fast oscillating irregularities are present over the entire energy range accessible to the LAT.
These irregularities are more likely to be flattened out by the degraded energy dispersion.

\begin{figure}

\includegraphics[width = .49 \linewidth]{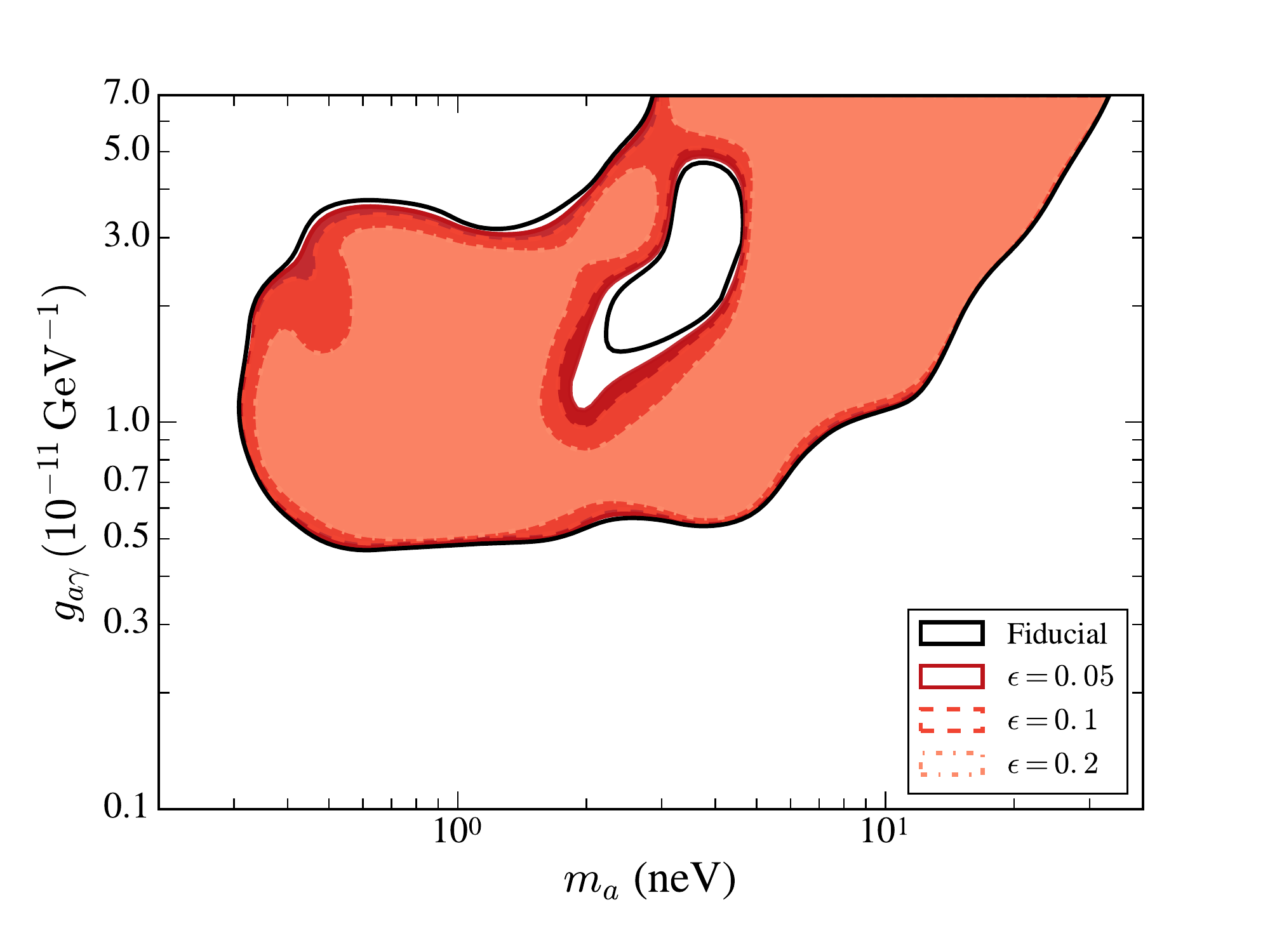}
\includegraphics[width = .49 \linewidth]{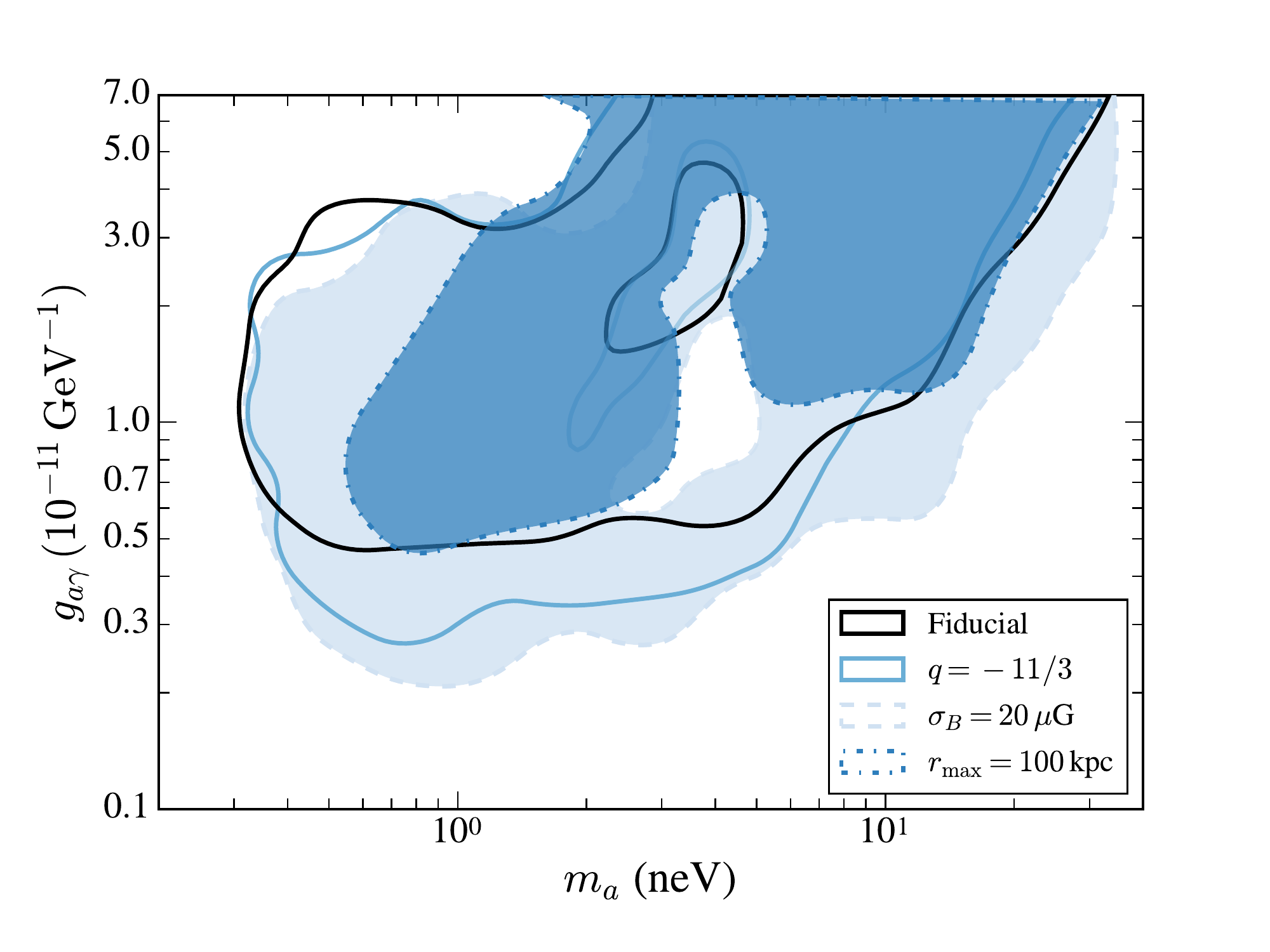}
\caption{\label{fig:sys}
\textit{Left}: Upper limits re-calculated for a degraded energy dispersion.
\textit{Right}: Upper limits re-calculated with different assumptions on the central magnetic-field strength ($\sigma_B = 20\,\mu$G),
the turbulence spectrum ($q = -11/3$), and the spatial extent of the magnetic field in the cluster ($r_\mathrm{max} = 100\,$kpc). 
}
\end{figure}

\subsubsection{Intra-cluster magnetic field}
The magnetic field of the Perseus cluster on scales of tens and hundreds of kpc 
is unknown due to the lack of Faraday rotation measurements \cite{taylor2006}. 
We therefore investigate how the limits change if we alter the 
fiducial assumptions for the $B$-field calculation. 

For fixed ALP parameters of $g_{11} = 1$ and $m_{a, \mathrm{neV}} = 3$,
Fig. \ref{fig:bparams} shows the difference in the likelihood $\Delta\ln\mathcal{L}$ 
that is obtained for different magnetic-field parameters. Higher values of $\Delta\ln\mathcal{L}$ correspond to stronger exclusions. 
In each panel, all other $B$-field parameters are fixed to their fiducial values. 
We choose again the random $B$-field realization that results in the $Q_B = 0.95$ quantile.
\begin{figure}
\includegraphics[width = 1 \linewidth]{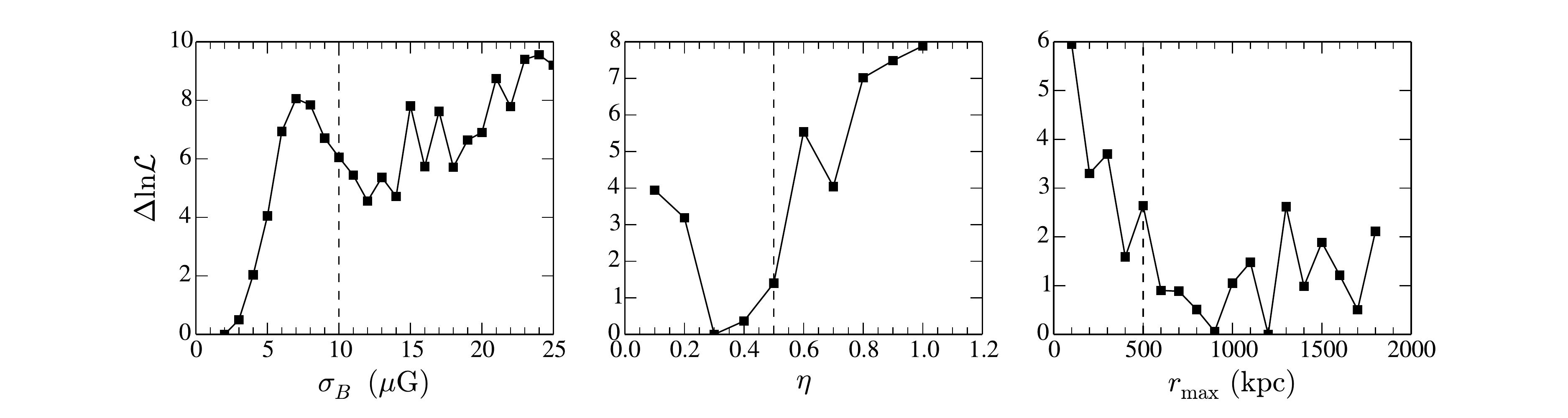}
\includegraphics[width = 1 \linewidth]{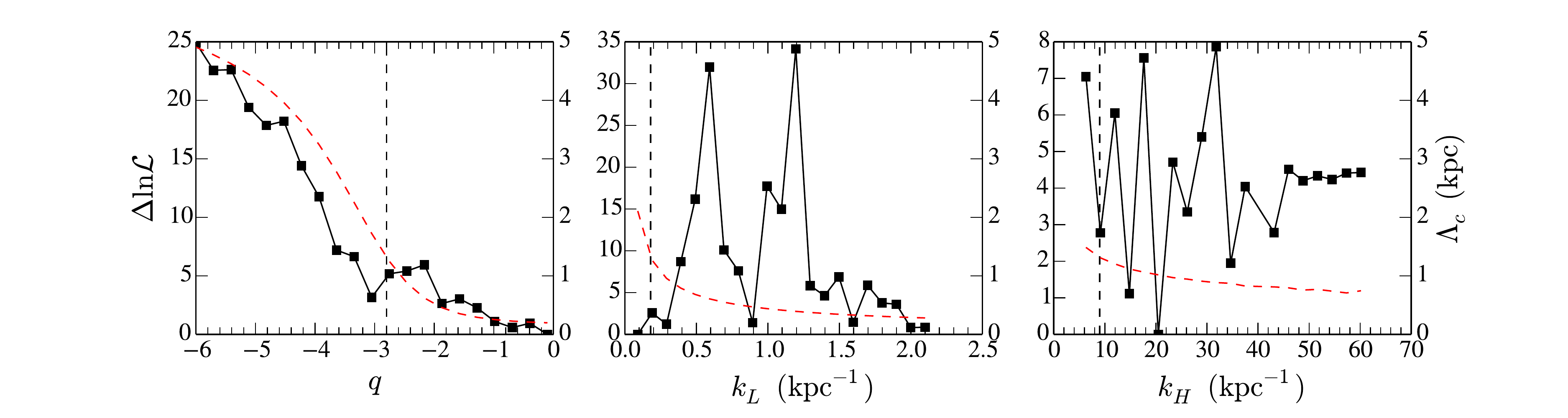}
\caption{\label{fig:bparams} 
Dependence of the maximum likelihood on the $B$-field parameters for fixed ALP parameters, 
$m_{a, \mathrm{neV}} = 3$ and $g_{11} = 1$. 
From left to right and top to bottom the tested parameters are the variance of the magnetic field ($\sigma_B$), 
the  index of the radial dependence of the $B$ field on the electron density ($B(r)\propto n_e(r)^\eta$),
the maximal spatial extent of the $B$ field ($r_\mathrm{max}$), the index of the turbulence spectrum ($q$), and 
the wave numbers corresponding to the maximum ($k_L$) and minimum ($k_H$) turbulence scales.
For the lower panel, the transversal coherence length  $\Lambda_c$ of the magnetic field 
is also shown (red dashed line and right $y$-axis). The vertical dashed lines denote the fiducial parameters.
}
\end{figure}
The top right panel shows the dependence on the central magnetic field. Increasing the $B$ field 
from 2\,$\mu$G to 7\,$\mu$G leads to a sharp increase in $\Delta\ln\mathcal{L}$. Above 7\,$\mu$G,
the photon-ALP mixing is maximal and changing the $B$-field changes 
the shape of the irregularities (but not their amplitude) leading to the fluctuations of  $\Delta\ln\mathcal{L}$.
In the top central panel, the dependence of the radial magnetic field on the electron distribution is increased,
where higher values of $\eta$ correspond to a stronger decrease of the $B$ field with radius. 
Surprisingly, the likelihood difference increases above $\eta = 0.5$. 
Inspecting the photon survival probability $P_{\gamma\gamma}$ one sees that the very rapid oscillations
 become suppressed for higher values of $\eta$ whereas larger features are more pronounced leading to stronger exclusions.
The situation is similar when one increases the distance from the cluster center up to where we assume a non-zero $B$ field (top left panel of Fig. \ref{fig:bparams}).
Naively one would assume that the considered ALP parameters should be more constrained for larger cluster radii. 
However, while the overall amplitude of the oscillation does not change, the frequency decreases with smaller radii
leading to potentially broader irregularities that are not washed out by the energy dispersion.

The lower row of Fig. \ref{fig:bparams} shows the dependence on the parameters of the turbulence spectrum. 
Changing these parameters will also change the coherence length of the transversal magnetic field 
$B_\perp$
 \cite{meyer2014},
\begin{equation}
\Lambda_c = \frac{1}{\langle B_\perp(\mathbf{x})B_\perp(\mathbf{x})\rangle}
\int\limits_0^\infty\mathrm{d}x_3 \langle B_\perp(\mathbf{x})B_\perp(\mathbf{x} + x_3\hat{\mathbf{e}}_3)\rangle,
\end{equation}
where $x_3$ is the axis along the line of sight. The fiducial parameter set results in a coherence length of $\Lambda_c \sim 1.32\,$kpc.
For softer turbulence spectra, i.e. smaller values of $q$,
the coherence length increases as large spatial scales become more dominant.  This leads 
to broader irregularities in energy, and, subsequently, to larger ALP exclusion areas. For $q \to 0$, the turbulence spectrum resembles white noise resulting 
in fast oscillations of $P_{\gamma\gamma}$ with energy.

The situation is less obvious when changing the minimum wave number, $k_L$ (or maximum turbulence scale). 
Even though $\Lambda_c$ slightly decreases with increasing $k_L$, 
strong maxima are visible in the likelihood difference (bottom central panel of Fig. \ref{fig:bparams}). 
These correspond to cases where $P_{\gamma\gamma}$ exhibits strong isolated peaks 
at the onset of the irregularities that are in strong tension with the data. 
These peaks are also seen in Fig. \ref{fig:pgg} between 100\,MeV and 1\,GeV, however not as isolated. 
Our fiducial parameter choices do not lead to such features and can be regarded as conservative in this respect.

Increasing the maximum wavenumber $k_H$ above its fiducial value has only a very small effect on the likelihood
and the coherence length. With increasing $k_H$, the frequency of the oscillations around the critical energy 
is also increased, yet, when folded with the energy dispersion, these features are not detectable anymore. 


In conclusion, the derived ALP limits depend sensitively on the assumed $B$-field parameters. 
We quantify the effect further by repeating the limit calculation for all ALP parameters  
but changing one $B$-field parameter at a time. 
The excluded regions are shown in Fig. \ref{fig:sys} (right).
We limit ourselves to $\sigma_B$, $q$, and $r_\mathrm{max}$. 
Increasing $\sigma_B$ to $20\,\mu$G allows us to exclude smaller values 
of the photon-ALP coupling thanks to the full degeneracy of $B$ and $g_{a\gamma}$.
The higher value of $\sigma_B$ is close to the measurement reported in \cite{taylor2006} of $25\,\mu$G. 
As a result, the excluded region of ALP parameters increases by $\upnum43\,\%$.
Similarly, assuming a Kolmogorov-type turbulence spectrum, $q = -11 / 3$, as observed e.g. in Hydra A \cite{kuchar2011}
and Coma \cite{bonafede2010}, increases the excluded parameter region by $\upnum12\,\%$.
In this case, the coherence length increases to $\Lambda_c = 2.89$\,kpc.
Decreasing the maximal distance up to which the magnetic field is non-zero 
to 100\,kpc leads to a decrease by $\upnum16\,\%$,
even though it might increase the significance of the exclusion for some ALP parameters as noted above.
\fi
\end{document}